\documentclass[twocolumn]{aastex62}

\usepackage{amsmath}
\usepackage{color}
\usepackage{graphicx}

\usepackage[normalem]{ulem}

\received{2019}
\revised{2020}
\accepted{2021}
\submitjournal{ApJ}

\shorttitle{GRB radio afterglows}
\shortauthors{Kangas et al.}

\begin{document}

\title{The late-time radio behavior of GRB afterglows: testing the standard model}


\correspondingauthor{Tuomas Kangas}
\email{tkangas@stsci.edu}

\author[0000-0002-5477-0217]{Tuomas Kangas}
\affil{Space Telescope Science Institute \\
3700 San Martin Drive, Baltimore, MD 21218, USA}

\author{Andrew S. Fruchter}
\affil{Space Telescope Science Institute \\
3700 San Martin Drive, Baltimore, MD 21218, USA}



\begin{abstract}
We examine a sample of 21 gamma-ray burst (GRB) afterglow light curves at radio frequencies, and compare them to the X-ray and/or optical properties of the afterglows and to the predictions of the standard jet/fireball model. Our sample includes every \textit{Swift} GRB with an X-ray light curve indicating a jet break and with a published radio light curve, as well as several other targets with observed X-ray or and/optical jet breaks. We examine the late-time decline of each burst, and attempt to fit an analytical model based on the standard GRB afterglow equations to each data set. We show that most of the events in our \textit{Swift} GRB sample are incompatible with the radio light curve behavior predicted by conventional afterglow theory. Many exhibit a late-time radio decline incompatible with the post-break X-ray or optical afterglow. Only one radio afterglow in this sample, at any time, shows the eventually expected decline of $\sim t^{-2}$, although two others show it in their mm light curve. Several others remain consistent with the standard model if such a decline began after the observations. The radio behavior alone does not, however, indicate whether a GRB can be fit by our modeling code. Indeed, several of the well-fit GRBs may only appear so due to a lack of multi-wavelength data. While a second source of emission can account for some of the anomalous radio behavior, our tests indicate this is often not the case unless the main jet component is simultaneously suppressed. 

\end{abstract}



\section{Introduction} \label{sec:intro}

It is widely thought that the afterglow emission of a gamma-ray burst (GRB) originates as synchrotron radiation from electrons accelerated in the shock resulting from the interaction of the GRB jet with the circumburst medium \citep[CBM; e.g.][]{pacz93,sari98,piran04}. This standard fireball model predicts that the afterglow emission from X-ray to radio behaves as a series of (smoothly connected) power laws of the form $f_{\nu} \propto t^{\alpha}$, where the index $\alpha$ depends on the index of the electron energy distribution $p$ and on the frequency in relation to the (evolving) breaks in the afterglow spectrum  \citep[e.g.][]{granotsari02}. These break frequencies are the self-absorption frequency $\nu_a$, the characteristic synchrotron frequency $\nu_m$ and the cooling frequency $\nu_c$. The relativistic nature of the jet and the resulting beaming of the emission from the jet is expected to cause an achromatic steepening of the light curve, known as a jet break, when the beaming angle becomes similar to the jet opening angle \citep{mrees99, rhoads99,sari99}.

However, this model has not been able to explain all features of GRB afterglows. In particular, it has difficulty with what appears to be a population of radio quiet GRBs \citep{hancock13,ronning17,ronning18}. It was argued that these bursts (roughly a third of all GRBs) do not simply lack an observed radio afterglow due to insufficiently deep observations, but are indeed \emph{intrinsically} radio-quiet. \citet{ronning18} further suggested that the two populations may have different progenitors: the radio-loud GRBs are typically longer and more luminous in $\gamma$-rays (especially at GeV energies where only radio-loud GRBs are detected), and the radio-loud bursts alone show an anti-correlation between prompt duration and redshift. Active galactic nuclei (AGN) exhibit a similar radio-quiet vs. radio-loud dichotomy \citep[][]{xu99}. \citet{chiaberge11} have suggested that radio-loud AGN require more massive and faster-spinning black holes than radio-quiet ones.

Another issue has been raised by several authors \citep{eichler05,gs09,ressler17,warren18}. A population of quasi-thermal electrons can exist in the jet in addition to the normally assumed power-law distribution, as the shock may only accelerate a fraction of the electrons (which may be as low as 0.01). The effect of this would be additional emission on top of the expected synchrotron spectrum -- which should dominate in the radio while leaving higher frequencies largely unaffected \citep[except in the hot electron model of][]{ressler17}. Meanwhile, $\nu_a$ would be increased by a factor of $\sim30$.

Yet another problem was presented by \citet{k20}, hereafter referred to as K20. They analyzed the late afterglows of GRBs 160509A and 160625B between X-ray and radio, and found no post-jet-break decline in the radio light curve even a factor of 10 or 20 later than the observed optical/X-ray break. For GRB 160625B, in particular, the radio break was not observed even at $\sim200$ days. Furthermore, the observed radio light curves were better described by a single power law ($\sim t^{-1}$) than by numerical afterglow models \citep{veerten12b} in the decline phase. The only scenario permitted by the standard jet model \citep[using equations from][]{rhoads99,granotsari02} that produces a comparable behavior is a pre-break decline above $\nu_m$.

A tendency of radio light curves of GRBs to be flatter than expected was noted earlier, in the pre-\textit{Swift} era, by \citet{frail04}, who noted that while the X-ray and optical slopes at late times were tightly correlated, \textit{radio} and optical slopes were not. Some late-time radio light curves were even argued to flatten over time, which the authors explained with a transition to the non-relativistic phase. \citet{frail04} also brought up the possibility of changes in shock microphysics. Both they and K20 speculated that one possible solution might be a two-component jet where a wider, less energetic cocoon surrounds the core of the jet and would be responsible for the radio emission \citep[suggested for GRB 030329 by][]{berger03,peng05}; see also \citet{lazzati05}. Therefore radio emission from the 'main' component, commonly assumed to be the dominant source, may need to be suppressed somehow. \citet{pk04} also included in their sample some GRBs with jet breaks, where the effect of the jet break was not seen in the radio. Considering various explanations for decoupling optical and radio decay, they disfavored the two-component scenario and instead considered a semirelativistic reverse shock in a continuous inflow of ejecta more promising. \citet{p05} examined a sample of GRBs through numerical modeling; a structured jet was able to improve the fit over the standard jet model for several bursts, but still did not fit the radio light curve well without severe scintillation effects, while delayed energy injection into a reverse shock was feasible for three bursts. \citet{oates07} noted that a two-component jet might resolve the discrepancy between X-ray and optical in GRB 050802, where a jet break was only seen in X-ray; while \citet{ub07} suggested the reverse shock dominating \textit{all} afterglow frequencies. 

Based on the aforementioned studies, our understanding of the radio emission of GRB afterglows seems incomplete. In this paper, we follow up on the work of K20, attempting to shed more light on this picture by examining the radio light curves of 21 GRBs with an observed light curve break in the optical or X-ray. We compare the radio behavior to that of the higher frequencies, in particular concentrating on the radio decline rate, the presence or absence of an observed jet break, and the evolution of the peak frequency in the radio spectra. We show that most of the events in our sample, at radio frequencies, do not behave as expected after the X-ray jet break. We also attempt to fit each burst using the standard jet model; yet, in many cases a single power law is a better description of the radio decline. In Section \ref{sec:analysis} we present our sample and the methods we use to examine the properties of these GRBs through power-law fits and analytical modeling. Our results are presented in Section \ref{sec:results}, and we discuss these findings in Section \ref{sec:disco} and finally summarize our conclusions in Section \ref{sec:concl}. We adopt the notation $F_{\nu} \propto t^\alpha \nu^\beta$ for power-law light curves and spectra.

\section{Data and analysis} \label{sec:analysis}

\subsection{The sample} \label{sec:sample}

\begin{table*}
\centering
\begin{minipage}{\linewidth}
\centering
    \begin{tabular}{lcccc}
       GRB & $z$ & $E_{\gamma,\mathrm{iso}}$ (erg) & $T_{90}$ (s) & Reference(s)\\
         \hline
       Group 1 \\
       \hline
        050820A & 2.615 & $8.9\times10^{52}$ & 241 (\textit{Swift}/BAT) & \citet{cenko06} \\
        051022 & 0.8\footnote{No optical afterglow was detected, but the GRB was localized to a likely host galaxy at $z\approx0.8$.} & $4.4 \times 10^{53}$ & $\approx 200$ (Konus-\textit{Wind}) & \citet{golenetskii05,rol07} \\
        070125 & 1.547 & $1.1 \times 10^{54}$ & $\approx 75$ (Konus-\textit{Wind}) & \citet{golenetskii07,chandra08} \\
        090313 & 3.375 & $3.4 \times 10^{52}$ & 83 (\textit{Swift}/BAT) & \citet{melandri10} \\
        110709B & ...\footnote{No optical afterglow was detected. No $E_{\mathrm{iso}}$ available.} & ... & 810 (\textit{Swift}/BAT) & \citet{zauderer13} \\
        120326A & 1.798 & $3.2\times10^{52}$ & 69 (\textit{Swift}/BAT) & \citet{laskar15} \\
        130907A & 1.238 & $3.0 \times 10^{54}$ & 364 (\textit{Swift}/BAT) & \citet{veres15} \\
        141121A & 1.469 & $8.0 \times 10^{52}$ & $\approx 1200$ (Konus-\textit{Wind}) & \citet{golenetskii14,cucchiara15} \\
        151027A & 0.81 & $4.0 \times10^{52}$ & 130 (\textit{Swift}/BAT) & \citet{nappo17} \\
        160509A & 1.17 & $8.6 \times 10^{53}$ & 371 (\textit{Fermi}/GBM) & \citet{roberts16,laskar16}; K20 \\
        160625B & 1.406 & $3.0 \times 10^{54}$ & 460 (\textit{Fermi}/GBM) & \citet{burns16,alexander17}; \\
         & & & & \citet{troja17}; K20 \\
        171010A & 0.33 & $2.2\times10^{53}$ & 104 (\textit{Fermi}/GBM) & \citet{poolakkil17,bright19} \\
        \hline
       Group 2 \\
       \hline
        990510 & 1.62 & $2.9\times10^{53}$ & 68 (\textit{BATSE}) & \citet{kippen99,veerten12b}\footnote{\citet{veerten12b} compiled their data from \citet{harrison99,israel99,bloom99,beuermann99,stanek99}, and \citet{pudalski99}.} \\
        991208 & 0.706 & $1.2\times10^{53}$ & 68 (\textit{Ulysses}) & \citet{hurley2000,galama00,galama03}; \\
         & & & & \citet{ctirado01} \\
        000301C & 2.04 & $5.4\times10^{52}$ & $\approx10$ (\textit{Ulysses}) & \citet{smith00,berger00,sagar00}; \\
         & & & & \citet{jensen01,rf01}\\
        000926 & 2.08 & $2.6\times10^{53}$ & $\approx25$ (\textit{Ulysses}) & \citet{hurley00,harrison01};\\
         & & & & \citet{price01}\\
        100418A & 0.624 & $1.0 \times 10^{51}$ & 8 (\textit{Swift}/BAT) & \citet{moin13,dup18} \\
        111215A & 2.06\footnote{No optical afterglow was detected -- photometric host redshift given by \citet{vdh15}.} & $1.4 \times 10^{53}$ & 374 (\textit{Swift}/BAT) & \citet{zauderer13,vdh15} \\
        140311A & 4.954 & $1.0 \times 10^{53}$ & 70 (\textit{Swift}/BAT) & \citet{laskar18a} \\
        140903A & 0.351 & $6.0 \times 10^{49}$ & 0.3 (\textit{Swift}/BAT) & \citet{troja16} \\
        161219B & 0.148 & $1.8 \times 10^{50}$ & 7 (\textit{Swift}/BAT) & \citet{laskar18c} \\
        \hline
    \end{tabular}
    \caption{GRB sample examined in this study. Group 1 refers to GRBs from our search of the entire \textit{Swift} GRB Catalogue, and is thus considered representative; while Group 2 is the result of a literature search for additional GRBs and not necessarily representative.}
    \label{tab:sample}
\end{minipage}
\end{table*}

The sample of GRBs we examine here was constructed as follows. In order to select only GRBs with probable jet breaks from the \emph{Swift} X-Ray Telescope (XRT) GRB Catalogue\footnote{\url{https://www.swift.ac.uk/xrt_live_cat/}}, we picked each burst (as of January 2020) with at least one break in the X-ray light curve; where the last reported power-law slope is $\alpha_X < -1.5$; and where the change in the power law between the last and second-to-last slope is $\Delta \alpha_X \leq -0.5$. These X-ray light curves were then visually inspected, and cases where the light curve behavior was clearly non-canonical until relatively late times (e.g. flaring activity with no clear power law segments), or where there were only a few X-ray data points, were removed. We then selected those bursts with published radio data in the literature. In the end, this left us with 12 GRBs. We refer to this subsample as Group 1.

In addition, we have included 9 other GRBs with published radio and X-ray/optical light curves and with a break in at least one frequency band that is consistent with a jet break (i.e. too steep to be the passage of $\nu_c$). Events such as GRB 130427A \citep{perley14,depasq16} with no observed jet breaks were not included. In particular, we included those bursts examined by \citet{pk04} with optical and/or X-ray light curves indicative of a jet break. This latter part of the sample, referred to as Group 2, is not necessarily a representative subsample of the full GRB population showing jet breaks, but is also included as additional examples of radio behavior.

Both subsamples and the main properties of the GRBs are summarized in Table \ref{tab:sample}. The duration, $T_{90}$, is obtained from the \textit{Swift}/BAT GRB Catalog\footnote{https://swift.gsfc.nasa.gov/results/batgrbcat/} unless otherwise specified. The sample includes a wide variety of GRBs including an ULGRB and a SGRB, three 'dark' GRBs with no detected optical afterglow, along with a broad range of redshifts ($0.1475 \leq z \leq 4.954$) and isotropic-equivalent energies ($6.0 \times 10^{49}~\mathrm{erg} \leq E_{\gamma,\mathrm{iso}} \leq 3.0 \times 10^{54}$ erg).

\subsection{Power-law fitting} \label{sec:plfit}

We have examined the radio and X-ray light curves of our sample of GRBs as follows. With the exception of GRBs 990510, 000926 and 140903A, the X-ray light curves were obtained from the \emph{Swift}-XRT Lightcurve Repository\footnote{http://www.swift.ac.uk/xrt\_curves/} and converted to flux densities at 5 keV using {\sc pimms}\footnote{http://cxc.harvard.edu/toolkit/pimms.jsp} and parameters from \emph{Swift} (which in the case of GRB 130907A included a time-variable photon index). For the pre-\textit{Swift}-era GRB 990510, X-ray fluxes were taken from \citet{kuulkers00} and converted to flux densities using the reported parameters. Fluxes of GRB 000926, also pre-\textit{Swift}, were reported at 0.5 and 3 keV by \citet{harrison01}. The late-time \emph{Chandra} data associated with GRB 140903A were reported as flux densities at 1 keV; therefore we also convert the \emph{Swift} X-ray data to 1 keV flux densities using parameters in \citet{troja16}. Radio data were obtained from the sources listed in Table \ref{tab:sample}. The light curves at the fitted frequencies were in some cases augmented by interpolating between nearby frequencies, or by scaling to a nearby frequency assuming a power law spectrum \citep{granotsari02}. We ignore any observed rise period of the light curve and any early features attributed to flares, plateaux or a reverse shock in the literature. To the light curve after these features (i.e. the decline only) we have attempted to fit a single power law of the form $f_{\nu} = f_{\nu,0} t^{\alpha}$ and a broken power law of the form 
\begin{equation}
f_{\nu} = f_{\nu,0}\Big[\Big(\frac{t}{t_{b}}\Big)^{-\omega \alpha_{1}} + \Big(\frac{t}{t_{b}}\Big)^{-\omega \alpha_{2}}\Big]^{-\frac{1}{\omega}} ,
\end{equation}
where $t_b$ is the break time (whether a true jet break or not) and $\omega$ is a parameter describing the sharpness of the break. We perform the fit using a fixed $\omega$ of 3 and 10 for each burst, and choose the fit with the smallest $\chi^2$ \citep[these were the values used by][and subsequently by K20]{liang07}. When a broken power-law fit is possible (i.e. at least five points to fit, as the function has four free parameters), we determine the presence or absence of a break in the light curve using an F-test for equality of variance between a single and a broken power law. We require an improvement at a $P_F>0.95$ level, where $P_F$ is the probability of a smaller difference in variance if the fits are equally good, to accept the break as robust and at $P_F>0.8$ to consider it ambiguous\footnote{The notation $P_F$ is used instead of the conventional $p$ to avoid confusion with the index of the electron energy distribution $p$.}. We use the fit to estimate $p$ in the X-ray using standard closure relations; the break time $t_j$ if applicable; and whether the break is consistent with the expected post-break slope of $\alpha_2 = -p$ from lateral expansion at the speed of sound \citep{rhoads99} or with a steepening by by $t^{-3/4}$ or $t^{-1/2}$ depending on the density profile of the CBM \citep[][]{mrees99,panmesz99}. The latter, which we refer to as the edge effect, is only due to the edge of the jet becoming visible after an initial pseudo-isotropic phase, with no lateral expansion.

In the radio we have used the frequencies with enough points for a fit after any observed rising phase and/or early features in the light curve dominated by a reverse shock \citep[seen in GRBs 050820A, 160509A, 160625B, and 161219B;][]{cenko06,laskar16,alexander17,laskar18c} -- this means at least four points, so that we can better determine how well a power law fit matches the apparent decline. We make an exception in the case of GRB 140903A, where there are only three points available but they cleanly fit a single power law. Tables \ref{tab:results1} and \ref{tab:results2} list the results of our fits to Group 1 and 2 objects, respectively. 

\subsection{Analytical modeling} \label{sec:models}

\begin{table*}
\centering    
\begin{footnotesize}
\begin{minipage}{0.95\linewidth}
    \renewcommand{\arraystretch}{0.83}
    \begin{tabular}{lccccccl}
    \centering
       GRB & Band & Decline & $t_\mathrm{start}$ & $\alpha_1$ & $\alpha_2$ & $t_b$ & Notes \\
        \hline
        050820A & 5 keV & BPL & 0.05 d & $-1.14 \pm 0.02$ & $-1.68 \pm 0.13$ & $6.7 \pm 1.8$ d & ...\\
         & $r$ & BPL & 0.06 d & $-0.87 \pm 0.02$ & $-1.66 \pm 0.18$ & 6.7 d & $t_b$ fixed \\
         & $i$ & BPL & 0.06 d & $-0.86 \pm 0.02$ & $-1.69 \pm 0.14$ & 6.7 d & $t_b$ fixed \\
         & 8.6 GHz & SPL & 8 d & $-0.66 \pm 0.25$ & ... & ... & ...\\
         \hline
        051022 & 5 keV & BPL & 0.1 d & $-1.42 \pm 0.05$ & $-2.50 \pm 0.27$ & $2.7 \pm 0.6$ d & Ambiguous break \\
         & 4.9 GHz & SPL? & 1 d & $-0.46 \pm 0.17$ & ... & ... & Large scatter\footnote{Attributed to scintillation by \citet{rol07}.} \\
         \hline
        070125 & 5 keV & BPL & 0.5 d & $-0.71 \pm 0.50$ & $-2.00 \pm 0.09$ & $1.1 \pm 0.3$ d & Large $\alpha_1$ error \\
        & $r$ & BPL & 1 d & $-0.05 \pm 0.09$ & $-1.88 \pm 0.04$ & $1.2 \pm 0.1$ d & ... \\
         & 22.5 GHz & SPL & 10 d & $-0.58 \pm 0.05$ & ... & ... & ... \\
         & 14.9 GHz & SPL & 10 d & $-0.54 \pm 0.23$ & ... & ... & ... \\
         & 8.5 GHz & BPL? & 17 d & $-0.33 \pm 0.11$ & $-1.12 \pm 0.32$ & $88 \pm 41$ d & Ambiguous break \\
         & 4.9 GHz & SPL & 27 d & $-0.29 \pm 0.06$ & ... & ... & ... \\
         \hline
        090313 & 5 keV & BPL & 0.3 d & $-1.04 \pm 0.14$ & $-2.23 \pm 0.17$ & $1.0 \pm 0.2$ d & ... \\
         & $R$ & SPL & 0.015 d & $-0.95\pm0.04$ & ... & ... & Host-dominated after X-ray break \\
         & $i$ & SPL & 0.01 d & $-0.88\pm0.03$ & ... & ... & Host-dominated after X-ray break \\
         & 16 GHz & SPL & 3 d & $-0.31 \pm 0.01$ & ... & ... & ... \\
         \hline
        110709B & 5 keV & BPL & 0.01 d & $-0.91 \pm 0.03$ & $-1.57 \pm 0.04$ & $0.65 \pm 0.06$ d & ... \\
         & 5.8 GHz & SPL & 6 d & $-0.70 \pm 0.10$ & ... & ... & ... \\
         \hline
        120326A & 5 keV & BPL & 0.5 d & $-1.23 \pm 0.19$ & $-2.50 \pm 0.26$ & $1.4 \pm 0.5$ d & Ambiguous break \\
         & $r$ & BPL & 0.5 d & $-0.95 \pm 0.06$ & $-2.52 \pm 0.21$ & $2.8 \pm 0.3$ d & Flattens at late times \\
         & 92.5 GHz & BPL? & ... & ... & ... & ... & Seems to break at $\lesssim10$~d \\
         & 24.5 GHz & SPL & 4 d & $-0.87 \pm 0.07$ & ... & ... & ... \\
         & 19.2 GHz & BPL & 4 d & $-0.60 \pm 0.07$ & $-1.91 \pm 0.99$ & $44.7 \pm 15.3$ d & ... \\
         & 5 GHz & BPL? & 4 d & $-0.43 \pm 0.18$ & $-1.23 \pm 1.09$ & $44.7$ d\footnote{Fixed at the clearer break time of the 19.2 GHz light curve to make a BPL fit possible.} & Some scatter; ambiguous break \\
         \hline 
        130907A & 5 keV & BPL & 0.01 d & $-1.35 \pm 0.02$ & $-2.20 \pm 0.03$ & $0.25 \pm 0.02$ d & ... \\
         & 15 GHz & SPL & 0.5 d & $-0.71 \pm 0.04$ & ... & ... & ... \\
         & 11 GHz & SPL & 3 d & $-0.88 \pm 0.28$ & ... & ... & Some scatter \\
         \hline
        141121A & 5 keV & BPL & 0.5 d & $-0.46 \pm 0.11$ & $-2.21 \pm 0.19$ & $3.8 \pm 0.5$ d & ... \\
         & 15 GHz & SPL & 3 d & $-0.61 \pm 0.08$ & ... & ... & Hints of the 7 GHz light curve shape \\
         & 7 GHz & SPL? & 11 d & $-0.45 \pm 0.27$ & ... & ... & Not a clear SPL \\
         & 5 GHz & SPL? & 6 d & $0.19 \pm 0.29$ & ... & ... & Not a clear SPL \\
         & 3 GHz & SPL? & 16 d & $0.88 \pm 0.50$ & ... & ... & Not a clear SPL \\
         \hline
        151027A & 5 keV & SPL & 0.1 d & $-1.65 \pm 0.02$ & ... & ... & ... \\
         & $R$ & BPL & 0.05 d & $-0.80 \pm 0.29$ & $-1.98 \pm 0.13$ & $0.5 \pm 0.2$~d & Flattens at late times \\
         & 5 GHz & BPL? & ... & ... & ... & ... & A hint of a break \\
          \hline
         160509A & 5 keV & BPL & 0.4 d & $-1.20 \pm 0.06$ & $-1.96 \pm 0.09$ & $3.7 \pm 0.8$ d & ... \\
         & 9 GHz & SPL & 10 d & $-0.92 \pm 0.13$ & ... & ... & ... \\
         & 6 GHz & SPL & 10 d & $-0.91 \pm 0.11$ & ... & ... & ... \\
         \hline
         160625B & 5 keV & BPL & 0.1 d & $-1.24 \pm 0.02$ & $-2.23 \pm 0.15$ & $22 \pm 4$ d & ... \\
         & 22 GHz & SPL & 10 d & $-0.75 \pm 0.12$ & ... & ... & ... \\
         & 6.1 GHz & SPL & 20 d & $-1.08 \pm 0.11$ & ... & ... & ... \\
         \hline
         171010A & 5 keV & BPL & 0.25 d & $-1.29 \pm 0.06$ & $-1.98 \pm 0.27$ & $3.8 \pm 1.6$ d & ... \\
         & $R$ & SPL & 0.75 d & $-1.13 \pm 0.24$ & ... & ... & Flattens at late times; $>2.5$~d ignored \\
         & 15.5 GHz & SPL & 3 d & $-1.12 \pm 0.05$ & ... & ... & ... \\
        \hline
    \end{tabular}
    \caption{Results of our single or broken power-law fits to the decline in the X-ray and radio light curves of Group 1 GRBs (and in the optical if it is relevant to the analysis), ignoring any rise in the light curve or features attributed to a reverse shock or late engine activity unless otherwise specified in the text; we thus only include points after $t_\mathrm{start}$ in the fit. In 'Decline', SPL corresponds to a single power law and BPL to a broken power law.}
    \label{tab:results1}
\end{minipage}
\end{footnotesize}
\end{table*}

\begin{table*}
\centering    
\begin{footnotesize}
\begin{minipage}{0.95\linewidth}
    \renewcommand{\arraystretch}{0.83}
    \begin{tabular}{lccccccl}
    \centering
       GRB & Band & Decline & $t_\mathrm{start}$ & $\alpha_1$ & $\alpha_2$ & $t_b$ & Notes \\
        \hline
        990510 & 5 keV & SPL & 0.3 d & $-1.43 \pm 0.07$ & ... & ... & ... \\
         & $V$ & BPL & 0.1 d & $-0.94 \pm 0.01$ & $-2.02 \pm 0.03$ & $1.3 \pm 0.1$ d & ... \\
         & $i$ & BPL & 0.6 d & $-1.16 \pm 0.04$ & $-1.81 \pm 0.05$ & $1.4 \pm 0.2$ d & ... \\
         & 8.6GHz & SPL & 3 d & $-0.62 \pm 0.10$ & ... & ... & ... \\
         \hline
        991208 & $R$ & SPL & 2 d & $-2.41 \pm 0.05$ & ... & ... & Jet break before observations \\
         & $I$ & SPL & 2 d & $-2.37 \pm 0.25$ & ... & ... & Jet break before observations \\
         & 8.5 GHz & SPL & 10 d & $-1.00 \pm 0.09$ & ... & ... & ... \\
         & 1.4 GHz & SPL & 7 d & $-0.78 \pm 0.18$ & ... & ... & ... \\
         \hline
        000301C & $B$ & BPL & 1.5 d & $-0.61 \pm 0.07$ & $-2.30 \pm 0.17$ & $3.8 \pm 0.4$ d & ... \\
         & $R$ & BPL & 1.7 d & $-0.64 \pm 0.05$ & $-2.60 \pm 0.10$ & $4.6 \pm 0.2$ d & ... \\
         & $K$ & BPL & 1.7 d & $-0.20 \pm 0.05$ & $-2.21 \pm 0.09$ & $3.6 \pm 0.1$ d & ... \\
         & 8.5 GHz & SPL? & 10 d & $-0.74 \pm 0.07$ & ... & ... & Hint of a break at $\sim100$~d \\
         & 4.9 GHz & SPL & 40 d & $-1.73 \pm 0.36$ & ... & ... & Large gap in light curve; possible break \\
         \hline
        000926 & 3 keV & SPL & 2 d & $-2.07 \pm 0.47$ & ... & ... & Jet break before observations \\
         & $B$ & BPL & 0.8 d & $-1.22 \pm 0.23$ & $-2.26 \pm 0.22$ & $1.6 \pm 0.3$ d & ... \\
         & $R$ & BPL & 0.8 d & $-1.64 \pm 0.04$ & $-2.28 \pm 0.07$ & $2.0 \pm 0.3$ d & ... \\
         & 8.5 GHz & SPL & 10 d & $-0.68 \pm 0.12$ & ... & ... & ... \\
         & 4.9 GHz & SPL & 10 d & $-0.56 \pm 0.10$ & ... & ... & ... \\
         \hline
        100418A & 5 keV & BPL & 0.5 d & $-1.01 \pm 0.08$ & $-1.85 \pm 0.20$ & $4.5 \pm 1.2$ d & ... \\
         & 90 GHz & SPL & 1 d & $-0.60 \pm 0.06$ & ... & ... & ... \\
         & 8.5 GHz & SPL & 38 d & $-1.05 \pm 0.21$ & ... & ... & ... \\
         & 5.0 GHz & SPL & 29 d & $-0.41 \pm 0.08$ & ... & ... & ... \\
         \hline
        111215A & 5 keV & SPL & 0.1 d & $-1.35 \pm 0.03$ & ... & ... & May break at the same time as 93 GHz \\
         & 93 GHz & BPL & 1 d & $-0.20 \pm 0.06$ & $-1.73 \pm 0.38$ & $15 \pm 3$ d & Ambiguous break \\
         & 19.1 GHz & SPL & 10 d & $-1.08 \pm 0.04$ & ... & ... & ... \\
         & 6.7 GHz & SPL & 15 d & $-0.80 \pm 0.09$ & ... & ... & ... \\
         & 4.9 GHz & SPL & 15 d & $-0.56 \pm 0.04$ & ... & ... & ... \\
         \hline
        140311A & 5 keV & BPL & 0.1 d & $-1.14 \pm 0.11$ & $-1.93 \pm 0.48$ & $1.3 \pm 1.1$ d & Ambiguous break \\
         & 24.5 GHz & SPL & 4 d & $-0.88 \pm 0.07$ & ... & ... & ... \\
         & 19.2 GHz & SPL & 4 d & $-0.80 \pm 0.09$ & ... & ... & ... \\
         & 13.5 GHz & SPL & 4 d & $-0.63 \pm 0.12$ & ... & ... & ... \\
         & 8.6 GHz & complex & ... & ... & ... & ... & Late-time rebrightening \\
         \hline
        140903A & 1 keV & BPL & 0.05 d & $-0.95 \pm 0.13$ & $-2.31 \pm 0.18$ & $0.7 \pm 0.2$ d & ... \\
         & 6.1 GHz & SPL & 2 d & $-0.64 \pm 0.03$ & ... & ... & ... \\
         \hline
         161219B & 5 keV & BPL & 0.05 d & $-0.80 \pm 0.01$ & $-1.64 \pm 0.11$ & $17 \pm 3$ d & ... \\
         & $r$ & SPL & 0.08 d & $-0.61 \pm 0.01$ & ... & ... & SN-dominated points ($>2.7$ d) ignored\\
         & 104 GHz & BPL & 1 d & $-0.48 \pm 0.05$ & $-1.47$ & ... & Too few points for proper BPL fit \\
         & 11 GHz & SPL & 15 d & $-0.76 \pm 0.02$ & ... & ... & ... \\
         & 5 GHz & SPL & 15 d & $-0.63 \pm 0.18$ & ... & ... & ... \\
        \hline
    \end{tabular}
    \caption{As Table \ref{tab:results1}, but for Group 2.}
    \label{tab:results2}
\end{minipage}
\end{footnotesize}
\end{table*}

\begin{table}
\centering
\begin{small}
\caption{Upper and lower limits of each free parameter in our model fits.}
\begin{tabular}{lcc}
\hline
Parameter & lower limit & upper limit \\
 \hline
 $E_\mathrm{K,iso}$ & $10^{49}$ erg & $10^{56}$ erg\\
 $p$ & 2.0 & 3.2 \\
 $n_0$ & $10^{-5}$ cm$^{-3}$ & $10^{4}$ cm$^{-3}$ \\
 $A_*$ & $10^{-5}$ & $10^{4}$ \\
 $\epsilon_e$ & $10^{-5}$ & $1$ \\
 $\epsilon_B$ & $10^{-6}$ & $1$ \\
 $\theta_0$ & $0.01$ rad & $0.5$ rad \\
 $A_V$ & $0$ & $5$ mag \\
 \hline
\end{tabular}
\label{tab:priors}
\end{small}
\end{table}

For each burst, we have also attempted to fit the standard afterglow model to the available data in order to check whether this model is able to reproduce the radio behavior. Even when no single (asymptotic) power-law segment predicted by the standard model matches the single or broken power-law fit, one can possibly obtain a good fit with a smooth transition from one slope to another, for example when break frequencies pass through the observed radio bands at certain times, and conversely, constraints imposed by other data may limit the applicability of the model even when the power law does seem compatible with it. We have developed a Python-based fitting code using the analytical representation of the evolution of a GRB synchrotron spectrum in \citet{granotsari02} at times before the jet break and when there is no lateral expansion. After the jet break we have used the analytical model of \citet{vdhthesis} \citep[their Table 2.10; based on][]{rhoads99} for the evolution of each break frequency and the normalization of the flux in the case of exponential lateral expansion. We also fit for the transition to non-relativistic expansion; the evolution of the break frequencies and normalizing flux in the non-relativistic phase is treated as described in Table 2.11 in \citet{vdhthesis}.

As the break frequencies $\nu_c$, $\nu_m$, and $\nu_a$ evolve differently with time, their order and thus the shape of the overall spectrum changes with time as well. \citet{granotsari02} break the time evolution of the spectrum into five discrete regimes, determined by the order of the break frequencies. When one break frequency passes below another, the regime changes. Regime 1 corresponds to $\nu_a < \nu_m < \nu_c$, regime 2 to $\nu_m < \nu_a < \nu_c$, and regime 5 to $\nu_a < \nu_c < \nu_m$, in which case an additional absorption frequency, $\nu_{ac}$, splits off from $\nu_a$. The time evolution of $\nu_{ac}$ in the lateral expansion and non-relativistic cases, not covered by \citet{granotsari02} or \citet{vdhthesis}, was determined for the sake of completeness using \citet{gps00}; see Appendix \ref{sec:app1} for details. 

We assume a $5\rightarrow1\rightarrow2$ evolution over time. The evolution of these regimes is determined by Equations (1) -- (9) of \citet{granotsari02}, while their Table 2 lists the parameters of each break frequency and the flux at that frequency in each regime. The break frequencies $\nu_a$ and $\nu_c$ are further affected by inverse Compton (IC) cooling. This lowers $\nu_c$ compared to the \citet{granotsari02} predictions by a factor of $(1 + Y)^2$, where in regime 5, $Y \approx \sqrt{\epsilon_e/\epsilon_B}$ when $\epsilon_e \gg \epsilon_B$ \citep{sariesin01}, thus prolonging regime 5. Simultaneously, $\nu_a$ is increased by a factor of $1+Y$. After slow cooling begins at $t_{5\rightarrow1}$, the evolution of $Y$ is approximated as a power law $Y \propto (t/t_{5\rightarrow1})^{(2 - p)/(8 - 2p)}$ in the ISM case, and as $Y \propto (t/t_{5\rightarrow1})^{(2 - p)/(4 - p)}$ in the wind case, as per Equations (3.7) and (B1) in \citet{sariesin01}. We also include IC emission, as this can in some cases affect the X-ray light curve. Furthermore, we test each object for signs of Klein-Nishina suppression of the IC cooling described in \citep{nakar09}, using their Eq. 64; the process and its results are described in Appendix \ref{sec:app2}.

To eliminate discontinuities in flux or break frequency across transitions from one spectral regime to another we use the following prescription. We compute the break frequencies and normalization flux of the spectrum from model parameters only at $t_{5\rightarrow1}$ calculated by equalizing $\nu_m$ and $\nu_c$ in regime 1, including the effect of IC cooling. Their evolution over time is then described with broken power laws, with each break corresponding to times of transition either from one regime to another; the jet break, with either lateral expansion \citep{rhoads99} or only the edge effect \citep{mrees99,panmesz99}; or a transition to the non-relativistic phase. Using an approach similar to that of \citet{laskar14}, the spectrum itself is computed as a sum of the spectra in each regime, weighted using a function $w_i$ for regime $i$:
\begin{align}
\begin{split}
w_5 = (1 + (t/t_{5\rightarrow1})^\eta)^{-1}~,& \\
w_1 = (1 + (t/t_{5\rightarrow1})^{-\eta})^{-1}& + (1 + (t/t_{1\rightarrow2})^\eta)^{-1} - 1~, \\
w_2 = (1 + (t/t_{1\rightarrow2})^{-\eta})^{-1}&~,
\end{split}
\end{align}
where $t_{i\rightarrow j}$ is the transition time from $i$ to $j$, and $\eta$ is a sharpness parameter that we fix at 3. As each break in the spectrum is soft, the resulting light curve has no sharp breaks. 

We have used the Markov chain Monte Carlo (MCMC) package {\sc emcee} \citep{emcee} to find the model parameters and their uncertainties through $\chi^2$ minimization. The free parameters are $p$, isotropic-equivalent kinetic energy $E_{\mathrm{K, iso}}$, the fractions of energy in electrons and magnetic fields $\epsilon_e$ and $\epsilon_B$, the jet opening angle $\theta_j$, and the density of the CBM $n_0$ (ISM) or $A_*$ (wind). The optical, ultraviolet and infrared fluxes were corrected for Galactic reddening using the \citet{dustmaps} dust maps and the \citet{cardelli89} extinction law; host galaxy extinction was corrected for using the \citet{pei92} extinction law, with $A_V$ as an additional free parameter in the fitting. For each parameter, we have used a simple top-hat prior (for parameters other than $p$ and $A_V$, a logarithmic top hat) with upper and lower limits listed in Table \ref{tab:priors}. The jet break time $t_j$ is determined by model parameters \citep{vdhthesis}:
\begin{align}
\begin{split}
t_{j,\mathrm{ISM}} =& \frac{1 + z}{2} \Big(\frac{\theta_j}{0.126}\Big)^\frac{8}{3} n_0^{-\frac{1}{3}} \Big(\frac{E_{\mathrm{K, iso}}}{10^{52}\mathrm{~erg}}\Big)^{\frac{1}{3}}~, \\
t_{j,\mathrm{wind}} =& \frac{1 + z}{2} \Big(\frac{\theta_j}{0.160}\Big)^4 A_*^{-1} \Big(\frac{E_{\mathrm{K, iso}}}{10^{52}\mathrm{~erg}}\Big)~.
\end{split}
\end{align}
At this point, the bulk Lorentz factor $\Gamma$ is estimated as $\Gamma \approx \theta_j^{-1}$, and its evolution treated as a power law whose slope depends on the CBM profile and lateral expansion \citep{kumarzhang15}: $\Gamma \propto t^{-1/2}$ with lateral expansion, $\Gamma \propto t^{-3/8}$ in an ISM-type CBM without lateral expansion or $\Gamma \propto t^{-1/4}$ in a wind-type CBM without lateral expansion. A good approximation of the transition to the non-relativistic phase is when $\Gamma \sim 1.3$ \citep{wkf98}, and we therefore place the transition there. We have separately run each fit with and without lateral expansion -- as numerical simulations \citep[e.g.][]{zhangmac09, granotpiran12} suggest that exponential lateral expansion is not relevant for most afterglows -- and with both constant-density and wind-type CBM for each target. Furthermore, we have run each of these with sharpness parameters 3 and 10 for the jet break, as with the power law fitting. Out of these, the best fit is then shown; unless otherwise mentioned, this is simply the fit with the best $\chi^2$. 

Points attributed to early reverse shock features, supernovae or host galaxy contamination in our source papers (Table \ref{tab:sample}) were discarded. A 15 per cent uncertainty floor was added to the data in all cases to avoid the fits being driven by a few exceptionally precise points. The luminosity distance of each object was obtained using the redshifts in Table \ref{tab:sample} -- with the exception of GRB 110709B, where the real redshift is unknown, so a fairly typical value of $z=2.0$ was assumed\footnote{The assumed value of the redshift will change some of the fit values \citep[see][]{zauderer13}, but for our purposes this is not relevant as long as one redshift results in compatibility with the standard model.}. We used the cosmological parameters in \citet{cosmology}: $H_0 = 69.6$ km s$^{-1}$ Mpc$^{-1}$; $\Omega_m = 0.286$; $\Omega_\Lambda = 0.714$. The best-fitting parameters and their uncertainties are listed in Table \ref{tab:models}, along with the jet break time $t_j$, $\nu_m(t_j)$ and geometry-corrected energy $E_\mathrm{K}$ using the best-fit parameters.

\section{Results} \label{sec:results}

\begin{figure}
\centering
\includegraphics[width=0.95\columnwidth]{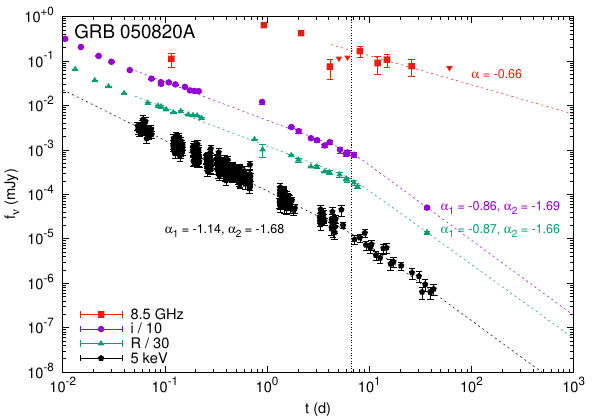} 
\includegraphics[width=0.95\columnwidth]{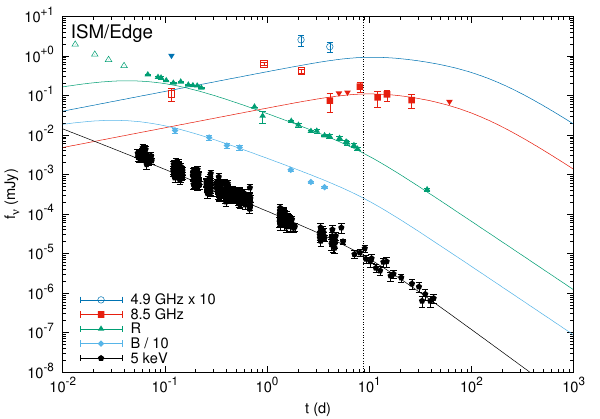}
\caption{Our single and broken power law fits (upper panel) and best-fitting analytical model (lower panel) compared to the light curves of GRB 050820A. Three additional optical frequencies used in the fitting are omitted from this figure for clarity.}
\label{fig:050820a}
\end{figure}

\begin{figure}
\centering
\includegraphics[width=0.95\columnwidth]{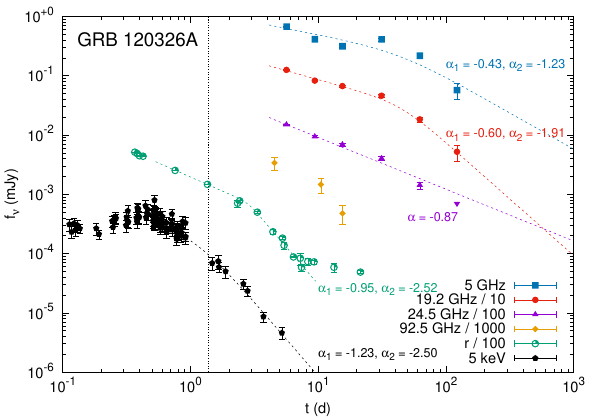}
\includegraphics[width=0.95\columnwidth]{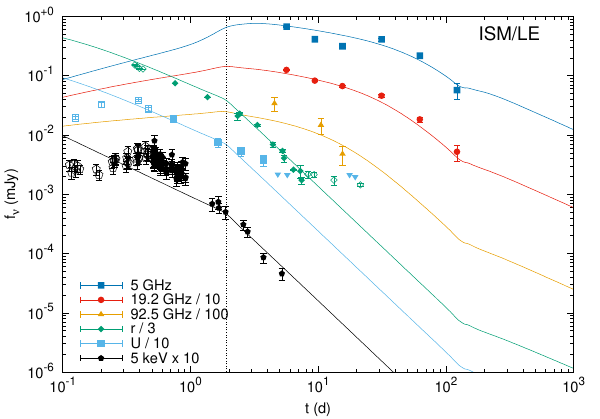}
\includegraphics[width=0.95\columnwidth]{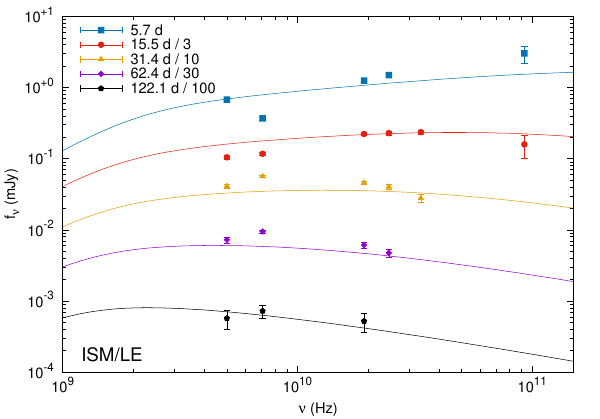}
\caption{Our single and broken power law fits to the light curves of GRB 120326A (upper panel), and our best-fitting analytical model compared to the light curves (middle panel) and radio SEDs (lower panel). Ten additional optical and radio frequencies used in the fitting are omitted from the light curve figure for clarity.}
\label{fig:120326a}
\end{figure}

\begin{figure}
\centering
\includegraphics[width=0.95\columnwidth]{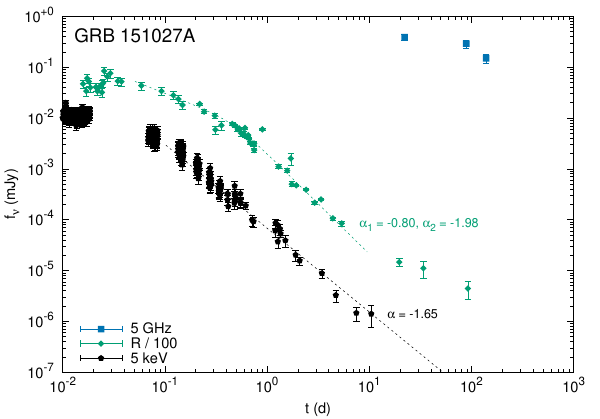}
\includegraphics[width=0.95\columnwidth]{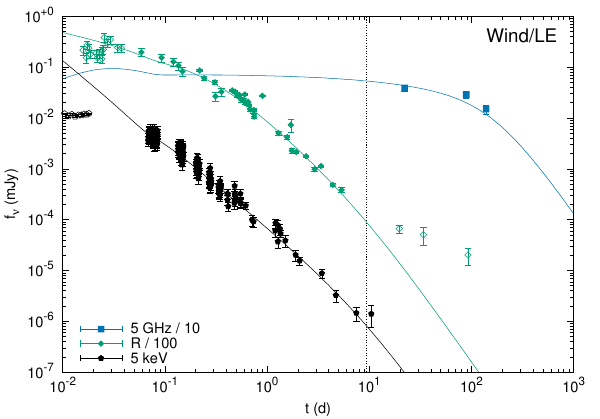}
\includegraphics[width=0.95\columnwidth]{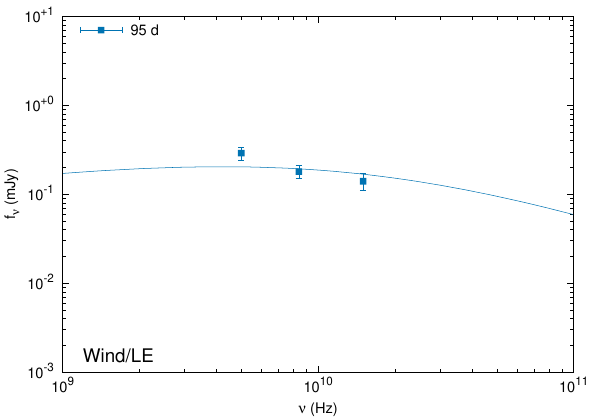}
\caption{Our single and broken power law fits to the light curves of GRB 151027A (upper panel), and our best-fitting analytical model compared to the light curves (middle panel) and radio SEDs (lower panel). Three additional radio frequencies with one data point each used in the fitting are omitted from the light curve figure for clarity.}
\label{fig:151027a}
\end{figure}

\begin{figure}
\centering
\includegraphics[width=0.95\columnwidth]{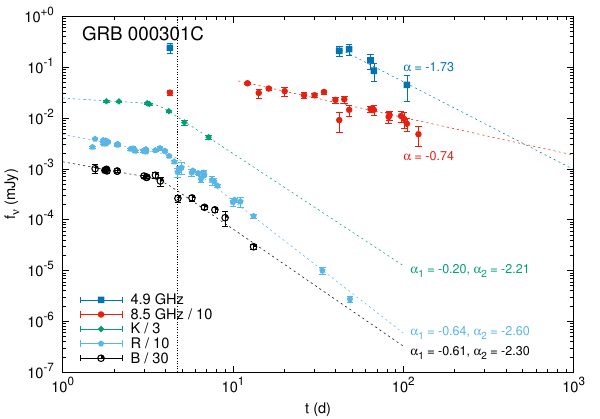} 
\includegraphics[width=0.95\columnwidth]{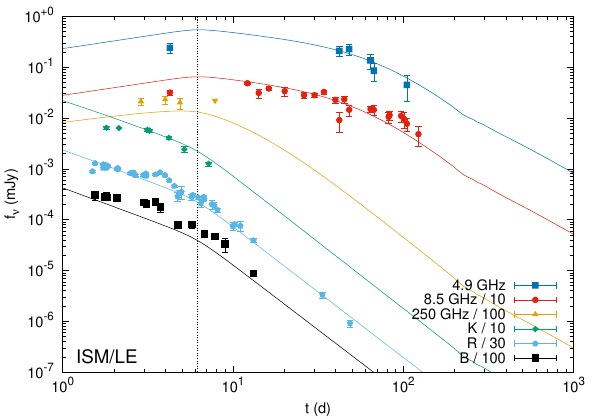}
\includegraphics[width=0.95\columnwidth]{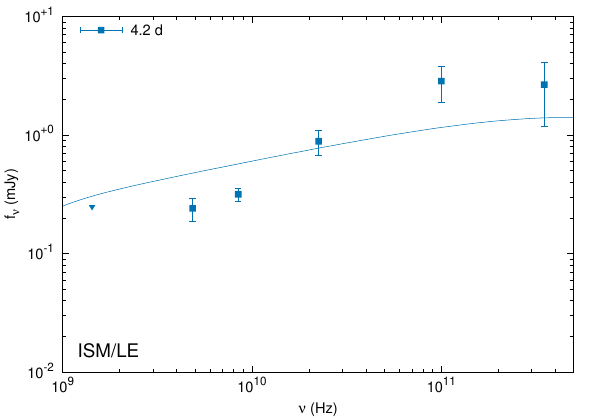}
\caption{Our single and broken power law fits (upper panel) to the light curves of GRB 000301C, and our best-fitting analytical model compared to the light curves (middle panel) and radio SEDs (lower panel). Ten additional optical and radio frequencies used in the fitting are omitted from the light curve figure for clarity.}
\label{fig:000301} 
\end{figure}

\begin{figure}
\centering
\includegraphics[width=0.95\columnwidth]{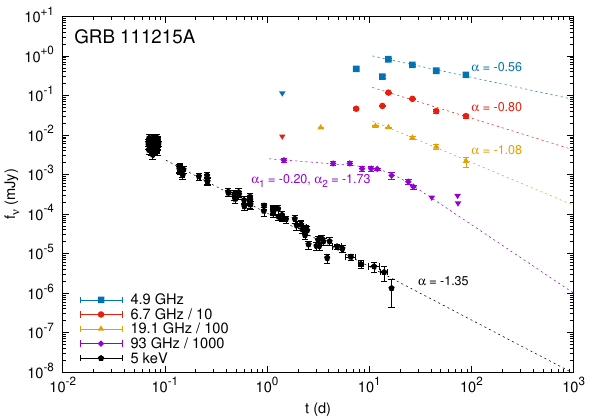}
\includegraphics[width=0.95\columnwidth]{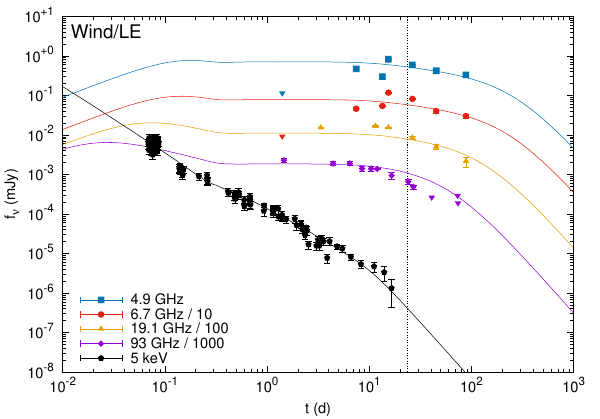}
\includegraphics[width=0.95\columnwidth]{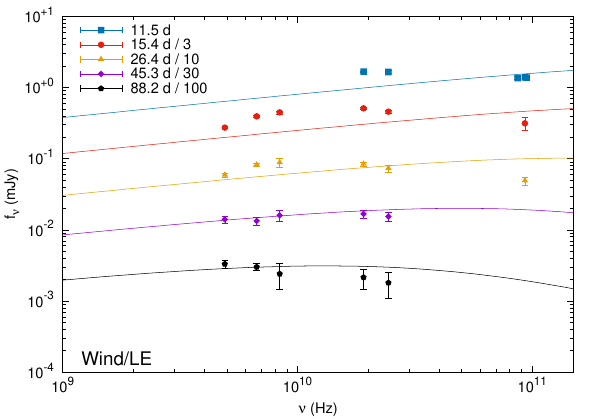}
\caption{Our single and broken power law fits to the light curves of GRB 111215A (upper panel), and our best-fitting analytical model compared to the light curves (middle panel) and radio SEDs (lower panel). Four additional radio frequencies used in the fitting are omitted from the light curve figure for clarity.}
\label{fig:111215a}
\end{figure}

Below, we describe the behavior of each individual GRB in our sample and whether some scenario of the conventional afterglow theory \citep[][]{rhoads99,granotsari02,vdhthesis} can account for all observations. For convenience, we summarize the scenarios and predicted light curve behavior in Table \ref{tab:stdmodel}. In Figs. \ref{fig:050820a} through \ref{fig:161219b} we show the single or broken power-law fits and, if available, the best-fit analytical model (with the CBM and jet break types of the best fit), compared to the observed light curves of each individual object in our sample. We also show the radio spectral energy distributions (SEDs) of the objects where light curves at multiple (at least 3) radio frequencies are available, compared to best-fitting model spectra. In each light curve, the dashed vertical line corresponds to the jet break time in either the X-ray broken power law fit or our best-fit model. If data are available at enough frequencies to make the figure cluttered, we do not show all available frequencies; instead, we present a selection of frequencies that adequately demonstrates the behavior of the GRB compared to the model. All available data in our sources (Table \ref{tab:sample}) are used for model fitting, apart from any epochs specified below for each object and marked with open symbols, e.g. because of early reverse shock features or host galaxy contamination.

\begin{table*}[p]
\centering
\rotatebox{90}{\begin{minipage}{1.36\linewidth}
\centering
    \begin{tabular}{lccccccccccc}
       GRB & Model & $E_\mathrm{K,iso}$ & $p$ & $n_0$ (cm$^{-3}$) & $\epsilon_e$ & $\epsilon_B$ & $\theta_j$ & $A_V$ & Best-fit $t_j$ & $\nu_m(t_j)$ & Best-fit $E_\mathrm{K}$ \\
       &  & ($10^{52}$ erg) & & or $A_{*}$ & &  & (rad) & (mag) & (d) & (Hz) & ($10^{52}$ erg) \\
         \hline
       Group 1 \\
       \hline
        \textit{050820A} & ISM/Edge & $408_{-63}^{+96}$ & $2.29\pm0.03$ & $1.4_{-0.3}^{+1.1} \times 10^{-5}$ & $0.06_{-0.01}^{+0.02}$ & $0.03\pm0.02$ & $0.027\pm0.002$ & $<0.01$ & 8.8 & $3.2\times10^{11}$ & $0.14$ \\
        \textit{051022} & Wind/Edge & $33.6_{-16.0}^{+49.0}$ & $2.57_{-0.12}^{+0.11}$ & $0.02_{-0.01}^{+0.03}$ & $0.13\pm0.06$ & $0.03_{-0.03}^{+0.11}$ & $0.03\pm0.01$ & ...\footnote{For dark GRBs (GRBs 051022, 110709B and 111215A), optical extinction is not relevant and was fixed at zero.} & 1.6 & $1.1\times10^{13}$ & $0.16$ \\
        070125 & ISM/LE & $9.2_{-0.4}^{+0.7}$ & $2.05_{-0.01}^{+0.02}$ & $7.3_{-1.1}^{+2.4}$ & $0.35\pm0.05$ & $0.43_{-0.16}^{+0.12}$ & $0.12\pm0.01$ & $0.07_{-0.02}^{+0.03}$ & 0.25 & $3.6\times10^{13}$ & $0.06$ \\
        090313 & ISM/Edge & $387_{-377}^{+277}$ & $2.04_{-0.03}^{+0.22}$ & $0.90_{-0.89}^{+0.24}$ & $0.04_{-0.01}^{+0.06}$ & $0.014_{-0.004}^{+0.005}$ & $0.04_{-0.01}^{+0.09}$ & $0.27_{-0.27}^{+0.14}$ & 0.8 & $1.2\times10^{11}$ & $0.33$ \\
        \textit{110709B} & Wind/LE & $35.8_{-12.6}^{+38.8}$ & $2.06_{-0.02}^{+0.03}$ & $0.11_{-0.05}^{+0.09}$ & $0.63_{-0.20}^{+0.21}$ & $0.032_{-0.028}^{+0.093}$ & $0.04\pm0.01$ & ... & 2.3 & $3.0\times10^{12}$ & $0.03$ \\
        120326A & ISM/LE & $15.0_{-1.2}^{+1.5}$ & $2.05\pm0.01$ & $0.40_{-0.06}^{+0.07}$ & $0.83_{-0.05}^{+0.04}$ & $0.15_{-0.04}^{+0.05}$ & $0.09\pm0.01$ & $0.35\pm0.02$ & 1.9 & $6.0\times10^{12}$ & $0.06$ \\
        130907A & Wind/LE & $441\pm23$ & $2.08\pm0.01$ & $0.03\pm0.01$ & $0.16\pm0.01$ & $8.5_{-1.1}^{+1.3} \times 10^{-3}$ & $0.010\pm0.001$ & 1.3\footnote{For GRB 130907A, extinction was fixed at $A_V = 1.3$ as per \citet{veres15} (see text).} & 0.25 & $1.8\times10^{13}$  & $0.02$ \\
        \textit{151027A} & Wind/LE & $72.1_{-14.6}^{+21.3}$ & $2.91_{-0.09}^{+0.07}$ & $0.03\pm0.01$ & $0.08_{-0.01}^{+0.02}$ & $0.07_{-0.04}^{+0.07}$ & $0.04\pm0.01$ & $1.1_{-0.2}^{+0.1}$ & 9.4 & $1.4\times10^{10}$ & $0.06$ \\
        \textit{160509A} & ISM/LE & $76.5_{-34.4}^{+131.7}$ & $2.08_{-0.03}^{+0.05}$ & $2.9_{-1.6}^{+2.6} \times 10^{-4}$ & $0.53_{-0.21}^{+0.27}$ & $0.012_{-0.011}^{+0.060}$ & $0.04\pm0.01$ & $2.9_{-0.4}^{+0.3}$ & 3.1 & $1.6\times10^{12}$ & $0.06$ \\
        160625B & ISM/Edge & $194_{-25}^{+31}$ & $2.38\pm0.02$ & $1.1_{-0.1}^{+0.2} \times 10^{-5}$ & $0.16\pm0.01$ & $0.03\pm0.01$ & $0.041\pm0.001$ & $0.07\pm0.02$ & 15.8 & $8.6\times10^{11}$ & $0.16$ \\
        171010A & ISM/LE & $10.5_{-2.4}^{+3.1}$ & $2.36\pm0.03$ & $8.9_{-0.7}^{+90.9} \times 10^{-5}$ & $0.06_{-0.02}^{+0.05}$ & $0.32_{-0.28}^{+0.38}$ & $0.06_{-0.01}^{+0.03}$ & $0.02_{-0.02}^{+0.04}$ & 4.0 & $4.2\times10^{11}$ & $0.02$ \\
        \hline
       Group 2 \\
       \hline
        \textit{990510} & Wind/LE & $28.7_{-10.7}^{+126.7}$ & $2.07_{-0.03}^{+0.13}$ & $0.06_{-0.03}^{+0.10}$ & $0.31_{-0.11}^{+0.16}$ & $0.03_{-0.03}^{+0.12}$ & $0.04\pm0.01$ & $<0.01$ & 1.6 & $1.6\times10^{12}$ & $0.02$ \\
        \textit{991208} & Wind/LE & $2.2\pm0.5$ & $2.09_{-0.01}^{+0.02}$ & $0.56\pm0.07$ & $0.83_{-0.10}^{+0.08}$ & $0.06_{-0.02}^{+0.03}$ & $0.12_{-0.01}^{+0.02}$ & $<0.01$ & 1.2 & $9.1\times10^{12}$ & $0.02$ \\
        000301C & ISM/LE & $88.9_{-69.6}^{+209.5}$ & $2.68_{-0.08}^{+0.05}$ & $1.4_{-1.0}^{+0.9}$ & $0.26_{-0.12}^{+0.13}$ & $7.4_{-7.0}^{+39.5} \times 10^{-5}$ & $0.13\pm0.02$ & $<0.01$ & 6.2 & $7.3\times10^{11}$  & $0.72$ \\
        \textit{000926} & ISM/Edge & $27.5_{-16.7}^{+5.0}$ & $2.27_{-0.08}^{+0.07}$ & $36.2_{-34.7}^{+3.8}$ & $0.29_{-0.18}^{+0.06}$ & $7.1_{-1.9}^{+128.9} \times 10^{-4}$ & $0.11_{-0.02}^{+0.01}$ & $0.10_{-0.02}^{+0.08}$ & 1.0 & $5.5\times10^{12}$ & $0.16$ \\
        100418A & ISM/Edge & $0.36_{-0.03}^{+0.11}$ & $2.26_{-0.02}^{+0.03}$ & $0.08_{-0.02}^{+0.03}$ & $0.38\pm0.03$ & $0.47_{-0.26}^{+0.12}$ & $0.45\pm0.02$ & $0.39_{-0.03}^{+0.04}$ & 38.6 & $8.3\times10^{10}$ & $0.03$ \\
        111215A & Wind/LE & $770_{-159}^{+201}$ & $2.92_{-0.08}^{+0.05}$ & $0.20\pm0.03$ & $0.12_{-0.02}^{+0.01}$ & $5.4_{-1.8}^{+4.0} \times 10^{-3}$ & $0.03\pm0.01$ & ... & 10.8 & $2.8\times10^{12}$ & $0.43$ \\
        140311A & ISM/LE & $2270_{-1860}^{+1380}$ & $2.21_{-0.08}^{+0.18}$ & $1.13_{-1.06}^{+0.80}$ & $0.37_{-0.23}^{+0.18}$ & $4.7_{-3.2}^{+249.1} \times 10^{-6}$ & $0.03_{-0.01}^{+0.03}$ & $0.69_{-0.08}^{+0.03}$ & 1.2 & $4.3\times10^{12}$ & $1.34$ \\
        \textit{140903A} & ISM/LE & $47.3_{-37.1}^{+244.0}$ & $2.46_{-0.08}^{+0.03}$ & $3.5_{-3.3}^{+246.8} \times 10^{-4}$ & $0.09_{-0.05}^{+0.07}$ & $1.57_{-1.53}^{+48.27} \times 10^{-4}$ & $0.03_{-0.02}^{+0.03}$ & $0.36_{-0.11}^{+0.05}$ & 1.1 & $4.1\times10^{11}$ & $0.03$ \\
        \hline
    \end{tabular}
    \caption{MCMC fit parameters and their uncertainties for each GRB in our sample. Both types of CBM (ISM and wind) and both types of jet break, with exponential lateral expansion (abbreviated as 'LE') and without it ('Edge'), were attempted for each burst, and the listed model is the best-fitting of these four. We note that in many cases, the best wind and ISM fits have roughly equal $\chi^2$. Names of GRBs that we consider consistent with the model are italicized.}
    \label{tab:models}
\end{minipage}}
\end{table*}

\begin{table*}[!]
\centering
\begin{minipage}{0.77\linewidth}
    \begin{tabular}{l|cc|cc|c|cc}
     \hline
    Frequency sequence & \multicolumn{2}{c|}{Pre-break} & \multicolumn{2}{c|}{Edge effect} & Lateral & \multicolumn{2}{c}{Non-relativistic} \\
    & $k=0$ & $k=2$ & $k=0$ & $k=2$ & expansion & $k=0$ & $k=2$ \\
     \hline
    $\nu < \nu_{ac} < \nu_a < \nu_c < \nu_m$ & $1/2$ & 1 & $-1/4$ & $1/2$ & 0 & $7/8$ & $-121/72$ \\
    $\nu_{ac} < \nu < \nu_a < \nu_c < \nu_m$ & $11/16$ & 1 & $-1/16$ & $1/2$ & $1/4$ & $7/8$ & $-133/72$ \\
    $\nu_{ac},\nu_a < \nu < \nu_c < \nu_m$ & $1/6$ & $-2/3$ & $-7/12$ & $-7/6$ & $-1$ & $2/3$ & $-2/3$ \\
    $\nu_{ac},\nu_a < \nu_c < \nu < \nu_m$ & $-1/4$ & $-1/4$ & $-1$ & $-3/4$ & $-1$ & $1/2$ & $1/6$\\
    $\nu_{ac},\nu_a < \nu_c < \nu_m < \nu$ & $-\frac{3p-2}{4}$ & $-\frac{3p-2}{4}$ & $-\frac{3p+1}{4}$ & $-\frac{3p}{4}$ & $-p$ & $-\frac{3p-4}{2}$ & $-\frac{7p-8}{6}$\\
     \hline
    $\nu < \nu_a < \nu_m < \nu_c$ & $1/2$ & 1 & $-1/4$ & $1/2$ & 0 & $-2/5$ & $2/3$\\
    $\nu_a < \nu < \nu_m < \nu_c$ & $1/2$ & 0 & $-1/4$ & $-1/2$ & $-1/3$ & $8/5$ & $4/9$\\
    $\nu_a < \nu_m < \nu < \nu_c$ & $-\frac{3(p-1)}{4}$ & $-\frac{3p-1}{4}$ & $-\frac{3p}{4}$ & $-\frac{3p+1}{4}$ & $-p$ & $-\frac{3(5p-7)}{10}$ & $-\frac{7p-5}{6}$\\
    $\nu_a < \nu_m < \nu_c < \nu$ & $-\frac{3p-2}{4}$ & $-\frac{3p-2}{4}$ & $-\frac{3p+1}{4}$ & $-\frac{3p}{4}$ & $-p$ & $-\frac{3p-4}{2}$ & $-\frac{7p-8}{6}$\\
     \hline
    $\nu < \nu_m < \nu_a < \nu_c$ & $1/2$ & 1 & $-1/4$ & $1/2$ & 0 & $-2/5$ & $2/3$\\
    $\nu_m < \nu < \nu_a < \nu_c$ & $5/4$ & $7/4$ & $1/2$ & $5/4$ & 1 & $11/10$ & $11/6$\\
    $\nu_m < \nu_a < \nu < \nu_c$ & $-\frac{3(p-1)}{4}$ & $-\frac{3p-1}{4}$ & $-\frac{3p}{4}$ & $-\frac{3p+1}{4}$ & $-p$ & $-\frac{3(5p-7)}{10}$ & $-\frac{7p-5}{6}$\\
    $\nu_m < \nu_a < \nu_c < \nu$ & $-\frac{3p-2}{4}$ & $-\frac{3p-2}{4}$ & $-\frac{3p+1}{4}$ & $-\frac{3p}{4}$ & $-p$ & $-\frac{3p-4}{2}$ & $-\frac{7p-8}{6}$\\
     \hline
    \end{tabular}
    \caption{Standard model predictions of the indices of power law segments in afterglow light curves. The values have been compiled using formulae in \citet{gps00}, \citet{granotsari02} and \citet{vdhthesis}; the edge effect values are the pre-break slopes plus a steepening by $t^{-3/4}$ or $t^{-1/2}$ when $k=0$ and $k=2$, respectively.}
    \label{tab:stdmodel}
    \end{minipage}
\end{table*}

\subsection{GRB 050820A}

The X-ray light curve is consistent with an edge effect jet break in wind-type CBM (the observed $\Delta \alpha = -0.54\pm0.14$); this in turn implies $p\approx2.2$. In an ISM scenario with the edge effect, the expected $\Delta \alpha = -3/4$ is $\sim1.6\sigma$ from the observed value. The optical decay is consistent with $p\approx2.2$ as well, but only in an ISM-type CBM and when $\nu_a, \nu_m < r,i < \nu_c <$~5 keV, as the X-ray decline at $<10$~d is steeper than the optical by $t^{0.27\pm0.03}$. \citet{cenko06} argued that the radio light curve includes a reverse shock before $\sim4$~d, although a reverse shock model was not explored in detail. The radio decay after this point can be fitted with a single power law; the slope of this power law is consistent with pre-jet-break expectations, assuming $p\approx2.2$, an ISM-type CBM and $\nu_m, \nu_a < 8.5~\mathrm{GHz} < \nu_c$; a post-break slope with edge effect and $\nu_a < 8.5~\mathrm{GHz} < \nu_m < \nu_c$ is within $\sim1.7\sigma$.

Our fitting process is done ignoring the radio reverse shock ($<4$~d). We also ignore the optical points before 0.05~d, as these seem to be the tail end of an achromatic flare visible all the way to the $\gamma$ rays \citep{cenko06}. Our best-fit model can account for all of the features of the light curve, but requires a very low ISM density ($\sim10^{-5}$~cm$^{-3}$) and a very small opening angle ($0.024$~rad, or $\sim1.4$~deg). If one restricts these parameters to more typical values (e.g. $\theta_j > 0.05$~rad and $n_0 > 10^{-4}$ cm$^{-3}$), the radio points are consistently over-predicted, while with wind CBM, they are under-predicted. We thus consider GRB 050820A tentatively consistent with standard theory. \citet{cenko06} argued that the radio emission of this burst was particularly weak and incompatible with the standard model, although they did not attempt a full multi-wavelength model fit. \citet{cenko10}, on the other hand, did, and consistently under-predicted the radio data instead. Their model \citep[from][]{yost03} used break frequencies and fluxes from \citet{sari98} and \citet{granot99a,granot99b} instead of the more recent \citet{granotsari02}, which may explain the discrepancy. 

\subsection{GRB 051022}
   
\begin{figure}
\centering
\includegraphics[width=0.95\columnwidth]{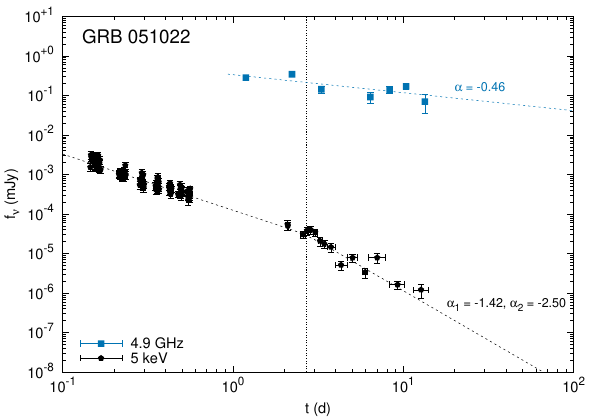} 
\includegraphics[width=0.95\columnwidth]{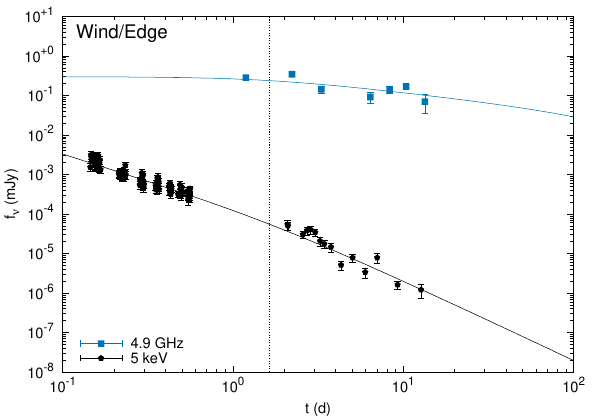}
\caption{Our single and broken power law fits (upper panel) and best-fitting analytical model (lower panel) compared to the light curves of GRB 051022.}
\label{fig:051022}
\end{figure}

A possible jet break was observed in the X-ray (with $P_F = 0.82$, the break is ambiguous); however, as this was a dark burst, an optical afterglow was not detected (Fig. \ref{fig:051022}). The X-ray light curve is consistent with $p\approx2.5$ and $\nu_c, \nu_m < \nu$. The radio light curve has a large scatter, but \citet{rol07} attribute this to scintillation effects; a single power law with strong scintillation may thus fit the light curve. In this case $\alpha_{\mathrm{radio}}$ is consistent with the $-1/3$ expected for a post-jet-break slope if $\nu < \nu_m$ and with full lateral expansion. Alternatively one can place a rise before $\sim2$ d, but then $\alpha_{\mathrm{radio}} = -0.67 \pm 0.28$. This is still consistent with $-1/3$ within $\sim 1.2 \sigma$, however, if $\nu_m$ stays above 4.9 GHz until $\sim15$ d. With an edge effect jet break ($1.1\sigma$ consistent with observations in the ISM scenario), the radio decline is also consistent with post-jet-break if $\nu_a < \mathrm{4.9~GHz} < \nu_c < \nu_m$. Our model fit adequately reproduces the light curve, though the best-fit $\theta_j = 0.03\pm0.01$ is somewhat low; the best fit has a wind-type CBM, but an ISM-type CBM produces a fit of comparable quality. Therefore GRB 051022 is consistent with the standard model. \citet{rol07} also achieve a good fit; their model does not include IC effects and assumes lateral expansion, but their parameters other than $p$ are consistent with ours.
    
\subsection{GRB 070125}
   
\begin{figure}
\centering
\includegraphics[width=0.95\columnwidth]{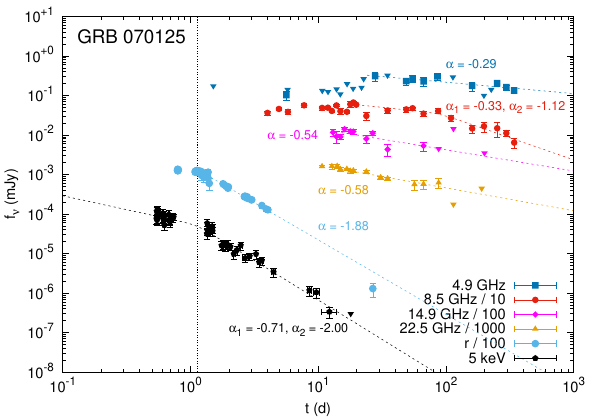}
\includegraphics[width=0.95\columnwidth]{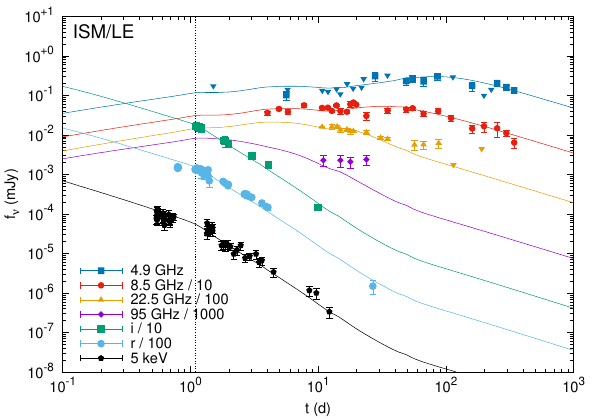}
\includegraphics[width=0.95\columnwidth]{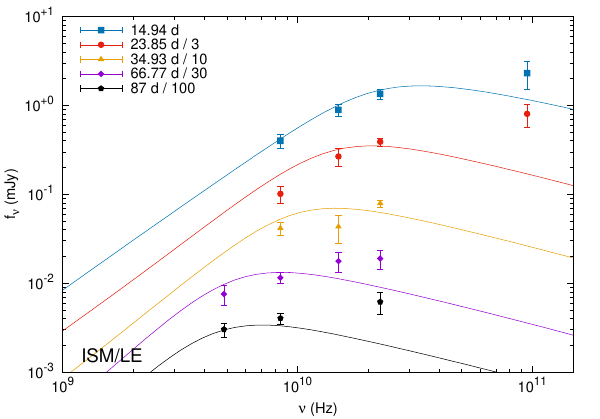}
\caption{Our single and broken power law fits (upper panel) to the light curves of GRB 070125, and our best-fitting analytical model compared to the light curves (middle panel) and radio SEDs (lower panel). Two additional optical frequencies used in the fitting are omitted from this figure for clarity.}
\label{fig:070125}
\end{figure}

\citet{chandra08} place an optical break $\gtrsim3\times$ earlier than the X-ray break, the latter delayed by effects from IC emission and not visible in the light curve. This is supported by the $r$-band light curve, which seems to steepen at $\sim3$~d, although the data are sparse here. However, if this is a jet break, the preceding slopes in $r$ ($-1.66\pm0.10$) and X-ray ($-2.00\pm0.09$) match an ISM-type CBM and $\nu_m, \nu_a < r < \nu_c < \mathrm{5~keV}$, but this requires $p\sim3.2$. If the jet break is in fact earlier and we have measured the post-jet-break slopes, this implies $p\sim2.3$ and no lateral expansion. The apparent plateau in the optical at $<1$~d may be due to $\nu_m$ passage in either case. 

The radio light curve (Fig. \ref{fig:070125}) shows consistency with a single power law at all fitted frequencies after a peak between $10$ and $\sim30$ d, but at 8.5 GHz we see an ambiguous ($P_F = 0.852$) break at $88 \pm 41$ d as well. The other frequency with an equally long follow-up, 4.9 GHz, does not share this feature, however, and the putative post-break slope is only $-1.12\pm0.32$. The radio slopes are inconsistent with scenarios allowed by the optical and X-ray bands. The spectral index below 22.5 GHz evolves from $1.23\pm0.11$ at 15~d to $0.47\pm0.04$ at 87 d, indicating that $\nu_a$ is located slightly above 22.5 GHz, but the slope of the light curve at 14.9 and 22.5 GHz before this time ($\alpha \approx -0.55$) does not match the $\alpha \geq -1/4$ predicted below $\nu_a$ after the jet break.

\citet{chandra08} note a lack of good radio fits to their model, but mostly bring up the early times, where they interpret the difference as scintillation effects. Their model, based on \citet{yost03}, under-predicts the radio data points at $>200$~d and, in the case of 8.46~GHz, also at $<10$~d. Our fitting code can roughly reproduce the X-ray, optical, 4.9 GHz and 8.5 GHz light curves \citep[with a jet break much earlier than suggested by][]{chandra08}, but under-predicts the late-time fluxes at 22.5 and 95 GHz. The shape of the SED clearly deviates from the model after 15 d. Therefore we consider GRB 070125 problematic for the standard model.
    
\subsection{GRB 090313} 

\begin{figure}
\centering
\includegraphics[width=0.95\columnwidth]{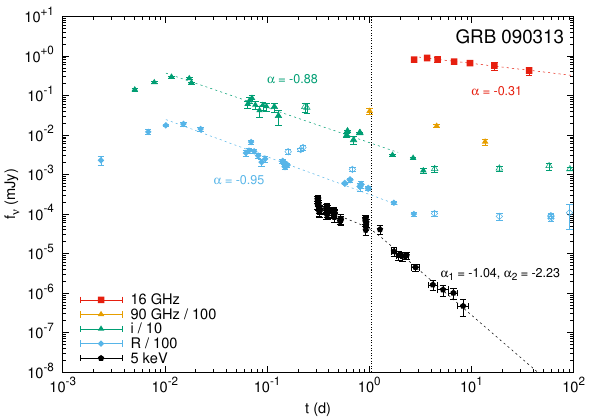} 
\includegraphics[width=0.95\columnwidth]{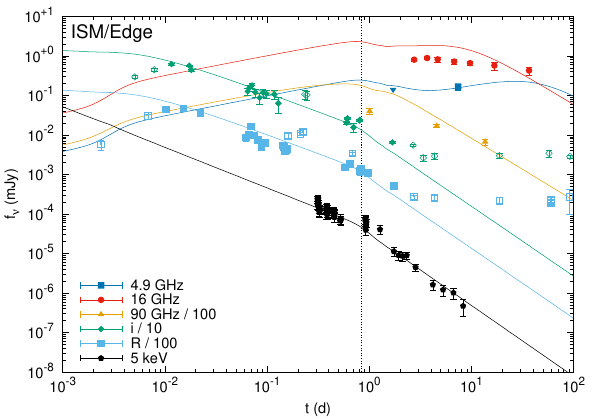}
\caption{Our single and broken power law fits (upper panel) and best-fitting analytical model (lower panel) compared to the light curves of GRB 090313. Four additional frequencies with one data point each, used in the fitting, are omitted from this figure for clarity.}
\label{fig:090313}
\end{figure}

The X-ray behavior is consistent with $p\approx2.2$ and $\nu_c < \nu$. The initial ($\lesssim0.01$~d) rise of the optical light curve can be explained by a decelerating fireball when $\nu_m$ lies below the optical bands \citep{melandri10}. The light curve also exhibits a rebrightening feature starting at $\sim0.2$~d. This 'bump' in the light curve may be due to a density variation in the CBM. Delayed energy injection is another option, but the post-injection optical light curve would then become much steeper than the X-ray decline. The radio light curve (Fig. \ref{fig:090313}) peaks around $3\times t_{j,X}$ and turns over onto a single power law with $\alpha_{\mathrm{16 GHz}} = -0.31 \pm 0.01$. This is close to $-1/3$ expected after a lateral expansion break when $\nu < \nu_m$, so GRB 090313 at first glance seems compatible with the standard model, if $\nu_m$ stays above 16 GHz until $\sim35$ d. The optical spectral index of this GRB, staying approximately constant around $-1.1$ \citep{melandri10}, indicates either that $\nu_m, \nu_c < i$ or that the optical data are affected by extinction.

\citet{melandri10} argue that the rise to the $\sim0.01$~d deceleration peak requires $\nu_m$ to be below optical at the start of the light curve. Our code does not include the deceleration peak, but we can address this by ignoring the points before 0.01~d while introducing an additional constraint: unless $\nu_m < R$ at $\sim0.003$~d, a zero likelihood for the model is returned. The optical bump at $0.2$--$0.5$~d was ignored in our model fits as well -- as were the points affected by the host galaxy after 2 d. In the fit, $\nu_m$ cannot stay above 16 GHz long enough to reproduce the observed shape of the light curve, and the model attempts to fit it with a smooth transition to regime 2. The radio points above 5 GHz are over-predicted until late times, where a steepening to a $t^{-p}$ decline is predicted, but not seen in the data. We do not consider GRB 090313 consistent with the standard model.
   
\subsection{GRB 110709B}

\begin{figure}
\centering
\includegraphics[width=0.95\columnwidth]{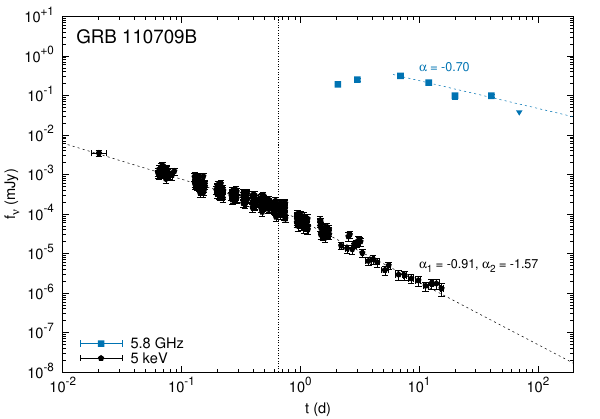} 
\includegraphics[width=0.95\columnwidth]{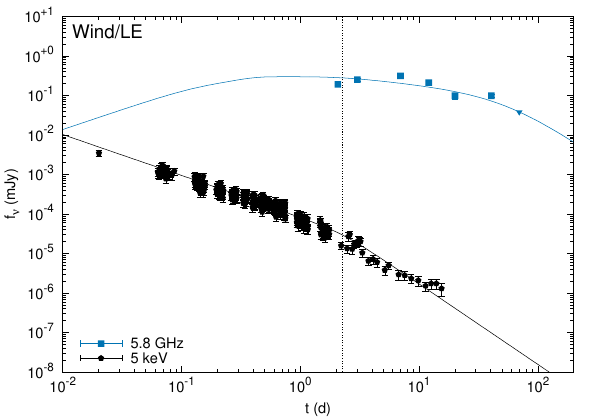}
\caption{Our single and broken power law fits (upper panel) and best-fitting analytical model (lower panel) compared to the light curves of GRB 110709B.}
\label{fig:110709b}
\end{figure}

GRB 110709B was a dark burst that was not detected in the optical. The X-ray behavior (Fig. \ref{fig:110709b}) is consistent with $p\approx1.9$ and $\nu_c, \nu_m < \nu$ or with $p\approx2.2$ and $\nu_m, \nu_a < \nu < \nu_c$ in ISM using standard closure relations. Thus we use Eqs. (4)--(7) of \citet{daicheng01} for $1 < p < 2$. Here, we find an X-ray post-jet-break slope consistent with $p\approx1.8$ and $\nu_c<\nu$ and an edge effect in ISM (albeit only at $1.8\sigma$), while in wind the observed $\Delta \alpha = 0.66\pm0.05$ is between lateral expansion and edge effect. The radio light curve at 5.8 GHz declines consistently with a pre-jet-break light curve when $p\approx1.8$, $\nu_m < 5.8~\mathrm{GHz}$ and in an ISM-type CBM, but if the slope is indeed pre-break in the radio, places a limit of $t_{j,\mathrm{radio}} \gtrsim 60t_{j,X}$ assuming that $t_{j,\mathrm{radio}}$ is after the last radio detection. These results do not change when using standard closure relations, except that $p\approx1.9$ instead. No post-jet-break scenario allowed by the X-ray light curve is consistent with the radio power law.

Both our fitting code and \citet{zauderer13} do, however, find a wind model that provides a reasonable fit to the radio if the post-jet-break light curve starts steepening around 50 d. The parameters of the fits are somewhat different, but \citet{zauderer13} did not include IC effects in their model. We find a roughly equally good fit with both types of CBM; our wind fit is shown as its $\chi^2$ is slightly better. GRB 110709B thus remains consistent with the standard model.

\subsection{GRB 120326A}

GRB 120326A is the only event in this sample with an apparent jet break clearly observed below $\sim100$~GHz; this happens at $\sim44.7\pm15.3$~d at 19.2 GHz, roughly 30 times later than the observed (albeit ambiguous; $P_F = 0.82$) X-ray jet break. The jet break is clearer in the optical. At 93 GHz the steepening occurs around 10 d. The post-break decline at all radio frequencies is highly uncertain, but consistent with the X-ray and optical decline. The X-ray and optical light curves imply $p\approx2.3$ and a lateral expansion break in an ISM-type CBM. Before the steepening at $\sim45$~d, the slope of the radio decline is seemingly inconsistent with theoretical expectations in this scenario.

This burst had a re-brightening episode before $\sim0.5$~d, for which \citet{laskar15} used a model with energy injection. As they point out, the behavior of the burst should be similar to the standard model without energy injection after this time. We therefore fit the light curves starting at 0.5~d. Our model reproduces the shapes of the radio (with a smooth transition from $t^{-1/3}$ to $t^{-p}$) and optical light curves, but the X-ray flux is under-predicted at $t<1$~d by a factor of a few and the steepening in the millimeter light curve is slightly later than observed. \citet{laskar15} obtained a nearly identical fit (apart from the non-relativistic transition) with similar parameters, and also under-predicted the X-ray flux. Both our best fit and \citet{laskar15} also place $\nu_m$ and $\nu_c$ below both optical and X-ray, but the power-law fit to the early X-ray is steeper than the optical by $t^{-0.28\pm0.20}$, implying that $\nu_m < r < \nu_c < \mathrm{5~keV}$. However, the optical and X-ray slopes are still within $1.4\sigma$. Nonetheless, the standard model seems to have a problem reproducing the afterglow after the energy injection.
 
\subsection{GRB 130907A}
   
\begin{figure}
\centering
\includegraphics[width=0.95\columnwidth]{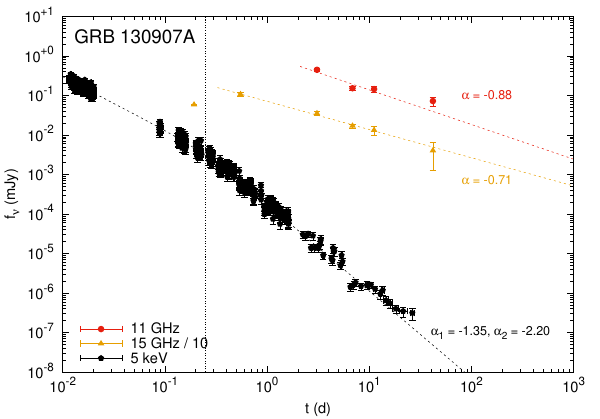}
\includegraphics[width=0.95\columnwidth]{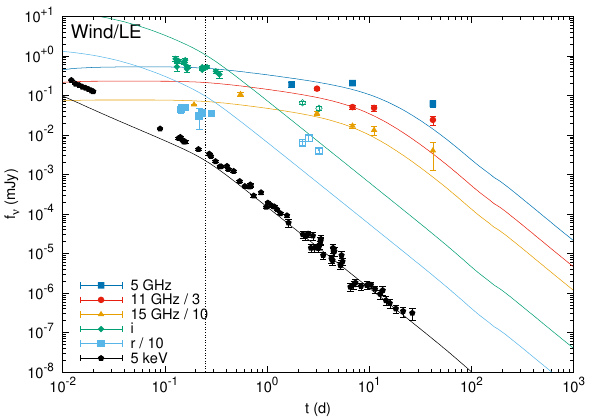}
\includegraphics[width=0.95\columnwidth]{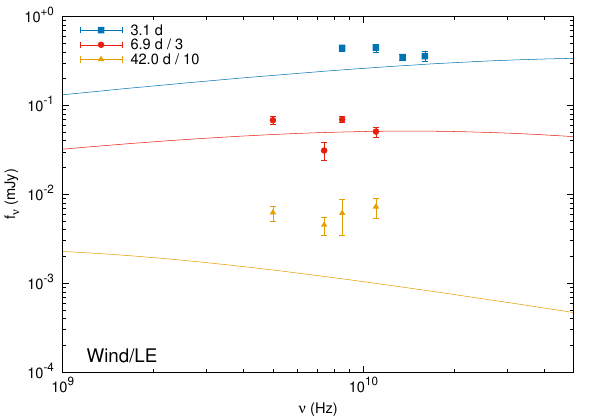}
\caption{Our single and broken power law fits to the light curves of GRB 130907A (upper panel), and our best-fitting analytical model compared to the light curves (middle panel) and radio SEDs (lower panel). Twelve additional optical and radio frequencies used in the fitting are omitted from the light curve figure for clarity.}
\label{fig:130907a}
\end{figure}

The X-ray light curve exhibits a jet break at $\sim0.2$~d (Fig. \ref{fig:130907a}). The pre-jet-break X-ray behavior is consistent with $p\approx2.4$ and $\nu_c < \nu$ in ISM, or $p\approx2.2$ and $\nu < \nu_c$ in wind. The X-ray spectrum has an index of $\beta_X =-0.69\pm0.06$ at 0.2 d \citep{veres15}, which implies $p=2.38\pm0.12$ -- after the jet break, on average, $\beta_X =-0.96\pm0.05$ implying a $\nu_c$ passage. In order to match the early X-ray-to-optical spectrum to a synchrotron model, one requires a host extinction of $A_V \approx 1.3$~mag and $\nu_m < i < \mathrm{5~keV} < \nu_c$ at 0.2 d \citep{veres15}. The post-jet-break slope is $\alpha_2 = -2.2\pm0.03$. Taken together, these observables seem to favor the wind scenario with lateral expansion and $p\approx2.2$.

The radio decline is a single power law inconsistent with the above scenario. \citet{veres15} described the radio evolution with a $-3(p-1)/4$ decline, which would result in $p\approx 2.0$ and, furthermore, requires an ISM-type CBM.
Since this refers to a pre-break slope, the radio break would also be delayed: $t_{j,\mathrm{radio}} \gtrsim 170 t_{j,\mathrm{X}}$ -- the longest relative delay in this sample. To account for this, \citet{veres15} attempted to explain the burst without a jet break and to describe the behavior at all frequencies as being pre-jet-break -- in this scenario the break in the X-ray light curve would be a combination of a $\nu_c$ passage and a transition from wind-like CBM to ISM-like. They were unable to reconcile this scenario with the steepness of the X-ray decline, though, and also suggested a combination of a narrow jet responsible for the X-ray emission and a wide jet responsible for the lower frequencies. The radio decline does match an edge effect break with wind and $\nu_a < \nu_c < 15~\mathrm{GHz} < \nu_m$, but this would not allow the 15 GHz light curve to rise after $t_j$ as observed.

For fitting, we fix extinction at $A_V = 1.3$, as the X-ray-to-optical spectrum seems to require it. Our fitting code is unable to fit any of the wavelengths well. A $t^{-p}$ decline is predicted in the radio, but not seen in the data. The deviation is driven by the radio data: a better fit to the optical and X-ray can be obtained by ignoring the early ($<10$~d) radio points, but even then only with $p>3$. Thus we consider this GRB inconsistent with the standard model.
    
\subsection{GRB 141121A}
   
\begin{figure}
\centering
\includegraphics[width=0.95\columnwidth]{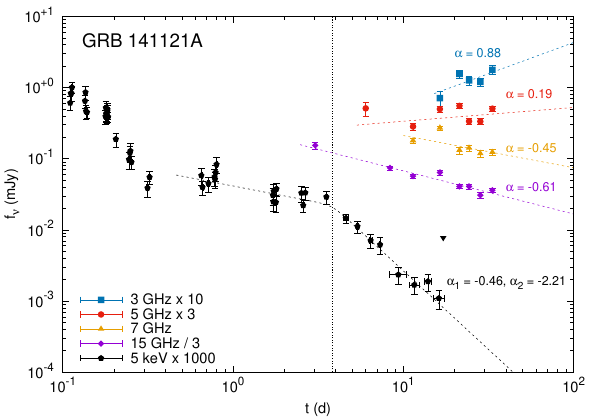}
\includegraphics[width=0.95\columnwidth]{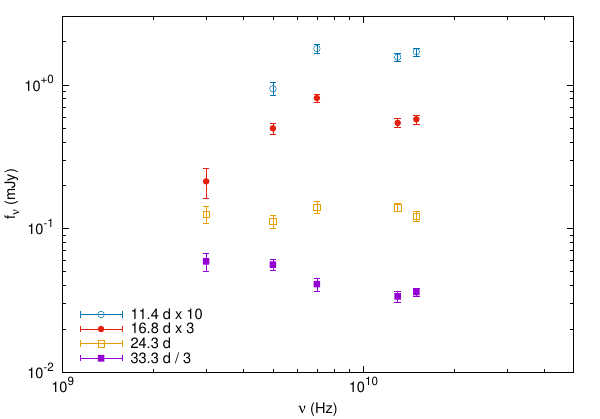}
\caption{Our single and broken power law fits to the light curves of GRB 141121A (upper panel) and the radio SEDs of this burst (lower panel).}
\label{fig:141121a}
\end{figure}

This ultra-long GRB shows peculiarities both in X-ray and in radio (Fig. \ref{fig:141121a}). The X-ray decline after the early steep decline ($\gtrsim0.4$~d) is a broken power law, and the decline after the last break at $\sim4$ d matches expectations for a post-jet-break decline with $p\approx2.2$ and full lateral expansion, or $p\gtrsim2.6$ without it. The slow decline before $\sim4$ d -- almost a plateau -- may be the product of late engine activity. The radio light curve at 15 GHz fits a single power law, consistent with an edge effect jet break, ISM and $\nu_a < \nu < \nu_c < \nu_m$ or with a pre-break slope, wind and $\nu_a < \nu < \nu_c < \nu_m$. The frequencies below 15 GHz show a more complex light curve, but its slope flattens and even becomes positive with decreasing frequency -- in this case, we also fit the rising light curves -- which also fits the pre-jet-break slope if 3 GHz $\lesssim \nu_a \lesssim 15$~GHz until $\sim 30$~d. In this case we can place a limit of $t_{j, \mathrm{radio}} > 8 \times t_{j, \mathrm{X}}$ based on the last radio detection. The 3 GHz rise is too steep for the post-jet-break scenario. The complex optical and radio light curve was interpreted by \citet{cucchiara15} using a model with energy injection into a reverse shock and a two-component jet. However, even this model does not fit the early radio points very well. The possible influence of a reverse shock is also seen in the spectrum, which shows some evidence of multiple peaks until 33.3 d. The complexity of the light curve and spectrum means our fitting code is naturally unable to reproduce the observations, and is therefore not used here. We consider GRB 141121A inconsistent with the standard model.

\subsection{GRB 151027A}

GRB 151027A exhibited a plateau in its early optical and X-ray light curve. The model favored by \citet{nappo17} incorporated a phase of late prompt emission to accommodate this. We ignore the plateau phase of the light curve ($\lesssim0.05$~d) in our fits; both the X-ray and optical light curves steepen after this. After 10~d, the optical light curve flattens, possibly because of supernova emission; this part is also ignored.

The X-ray light curve declines steeply enough to match our sample criteria; but only the optical light curve has a feature that looks like a jet break. After 0.5~d, the optical slope is steeper than the X-ray by $t^{-0.33\pm0.14}$, which is consistent with $\nu_m < R < \nu_c < \mathrm{5~keV}$ in a wind-type CBM. The slope of the optical light curve before this is inconsistent with this scenario, however.

Our best model fit to the post-plateau light curves reproduces the X-ray, optical and radio data. With a wind model, the steep decay is reproduced with no jet break until the end of the observed light curve. It does require a high $p$, however: $p = 2.91_{-0.09}^{+0.07}$. Thus we consider GRB 151027A tentatively consistent with the standard model. The shape of the X-ray-to-optical spectrum, which \citet{nappo17} considered a sign of the late prompt component, is addressed in our model by host galaxy extinction. Their late prompt emission model is, of course, also plausible as this extra component and the early plateau are not included in our code.

\subsection{GRB 160509A}
   
\begin{figure}
\centering
\includegraphics[width=0.95\columnwidth]{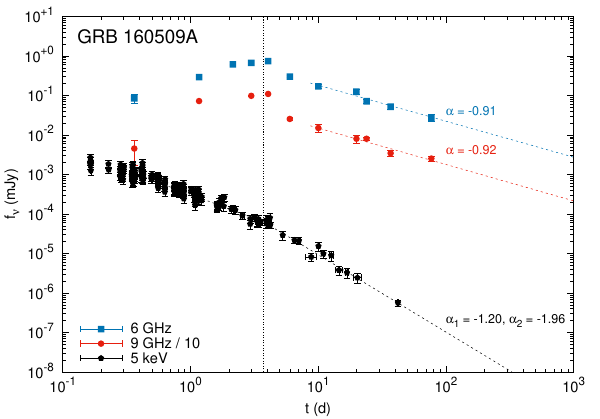}
\includegraphics[width=0.95\columnwidth]{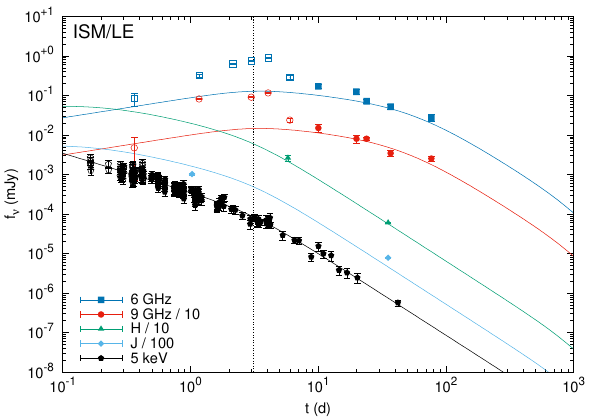}
\includegraphics[width=0.95\columnwidth]{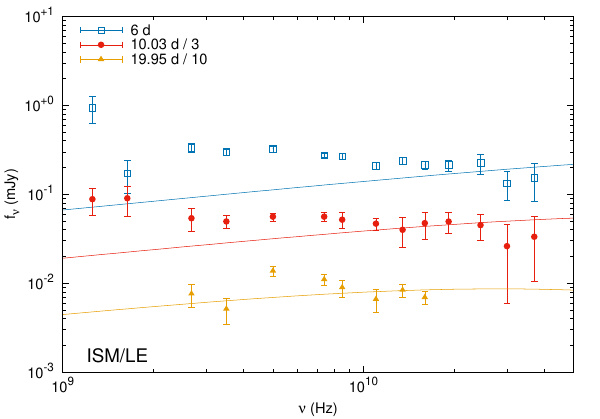}
\caption{Our single and broken power law fits to the light curves of GRB 160509A (upper panel), and our best-fitting analytical model compared to the light curves (middle panel) and radio SEDs (lower panel). Five additional optical and radio frequencies used in the fitting are omitted from the light curve figure for clarity.}
\label{fig:160509a}
\end{figure}

This burst was well observed in the X-ray, but an optical break was not observed due to a high extinction (although K20 did see a late-time slope consistent with post-jet-break expectations; see Fig.\ref{fig:160509a}). The radio light curves at 6 and 9 GHz (with the longest follow-up) are consistent with a pre-jet-break slope assuming $6~\mathrm{GHz} > \nu_m$ in an ISM-type CBM and $p \approx 2.2$, which is also consistent with the X-ray. The break time in X-rays is $\sim3.5$ d (K20), resulting in a limit of $t_{j, \mathrm{radio}} > 20 \times t_{j, \mathrm{X}}$ based on the last radio detection if the slope is indeed pre-break. Alternatively, the radio slope is consistent with ISM, edge effect and $\nu_a < \nu_c < \nu < \nu_m$, which the sparse optical data do not rule out, but this would place $\nu_c$ below 6 GHz at 10 d, which does not fit the radio-to-optical spectrum with $\beta = -0.40\pm0.01$ (K20). A reverse shock was detected in this GRB \citep{laskar16}. The spectrum initially shows a moving peak associated with the reverse shock, but after its passing ($\gtrsim10$~d based on the light curves) the spectrum becomes almost flat. This could in principle be due to a combination of reverse shock contribution and a smooth $\nu_m$ or $\nu_c$ break around the observed frequencies.

Our best-fit model with an ISM-like CBM does reproduce a flat radio SED at late times, and fits the light curves at all frequencies reasonably well. This requires a smooth $\nu_m$ passage through the radio bands between 10 and 100 days in order to be consistent with the observed decline. Thus, provided that the $\sim t^{-2}$ decline in the radio starts around the last observed epoch, and the flat spectral shape around 10~d can be attributed to lingering contribution from the reverse shock, GRB 160509A is consistent with the standard model. The parameters of our best fit differ somewhat from \citet{laskar16}, but using their parameters also results in a reasonable fit by eye, and they lacked access to the late-time data in K20.

We note that K20 used the numerical fitting code {\sc boxfit} \citep[][]{veerten12b}, and their best fit was only consistent with the late-time points ($\gtrsim30$ d), under-predicting the radio fluxes until then. Furthermore, even this required a jet break $\sim10$ times later than the broken power law fit indicates. This suggests that either the analytical or numerical model (or both) is erroneous. A further difference between the models was the extremely late jet break in the best-fit wind model -- and in a {\sc boxfit} light curve produced using the best wind parameters of \citet{laskar16} (neither of which fit the radio light curve). An ISM model with their parameters produced a similar result as the best fit. 
    
\subsection{GRB 160625B}
   
\begin{figure}
\centering
\includegraphics[width=0.95\columnwidth]{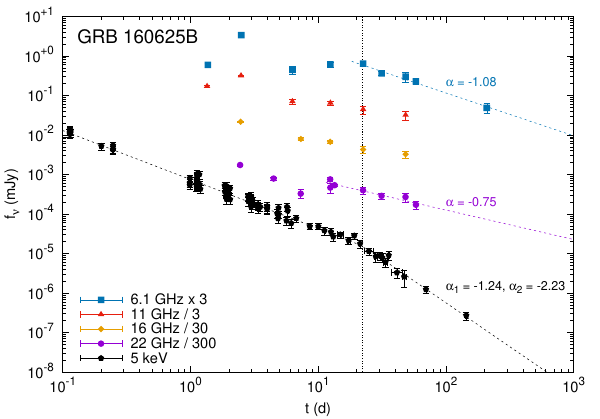}
\includegraphics[width=0.95\columnwidth]{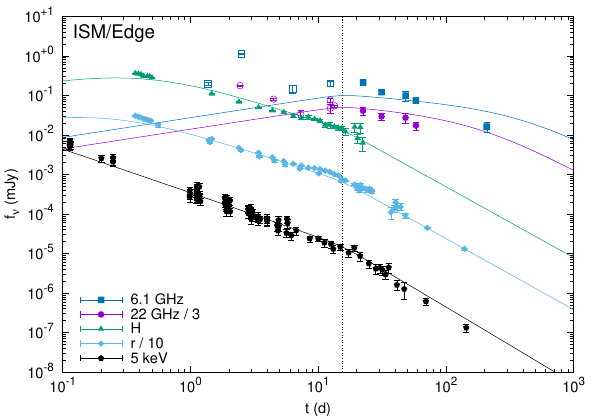}
\includegraphics[width=0.95\columnwidth]{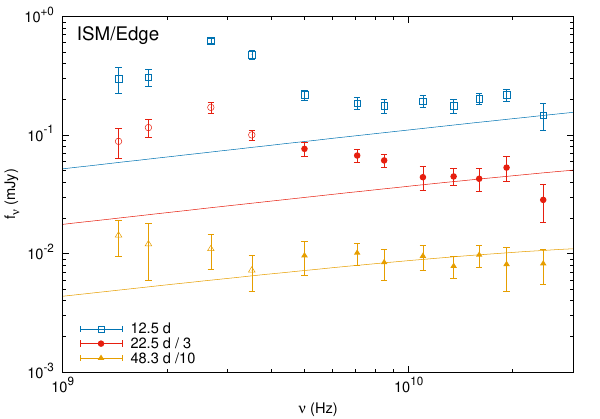}
\caption{Our single and broken power law fits to the light curves of GRB 160625B (upper panel), and our best-fitting analytical model compared to the light curves (middle panel) and radio SEDs (lower panel). Eleven additional optical and radio frequencies used in the fitting are omitted from the light curve figure for clarity.}
\label{fig:160625b}
\end{figure}

An optical and X-ray jet break was seen at $\sim20$ d (K20). The radio light curve (Fig. \ref{fig:160625b}) at 6.1 GHz is again consistent with a pre-break slope assuming $\nu_m < \nu$ in an ISM-type CBM, with a slope corresponding to $p \approx 2.4$, which is close to the value of $p\approx2.3$ determined through the early X-ray and optical light curve. Thus a limit of $t_{j, \mathrm{radio}} > 10 \times t_{j, \mathrm{optical}}$ can be placed if the pre-break scenario holds. No post-break slope allowed by the higher frequencies -- which imply ISM, $p\approx2.3$ and $\nu_m, \nu_a < \nu_c$ -- is consistent with the radio. This burst also showed signs of a reverse shock \citep{alexander17}, at low frequencies ($\lesssim5$ GHz) until $\sim20$ d. After this, similarly to GRBs 130907A and 160509A, the radio SED becomes flat. 

In our fitting, we ignore radio points at $<20$~d because of the reverse shock. Since \citet{alexander17} point out a possible extreme scattering event affecting the lowest radio frequencies, we also ignore all $<5$~GHz data. Our best-fit model is able to reproduce the observed behavior at optical and X-ray frequencies. However, the shapes of the 6.1 GHz light curve and the SED deviate somewhat from the model, the former both at early times ($\sim20$~d) and at the last radio point ($\sim200$~d). Furthermore, the model once again predicts a $\sim t^{-2}$ decline starting soon after the last observed radio epoch. Therefore, considering the problems K20 had with fitting the light curve using {\sc boxfit} as well, we consider GRB 160625B inconsistent with the model. \citet{alexander17} and \citet{troja17} found best-fit parameters similar to ours, but did not have the $210$~d data point and placed a steepening at $\sim50$~d.
    
\subsection{GRB 171010A}
   
\begin{figure}
\centering
\includegraphics[width=0.95\columnwidth]{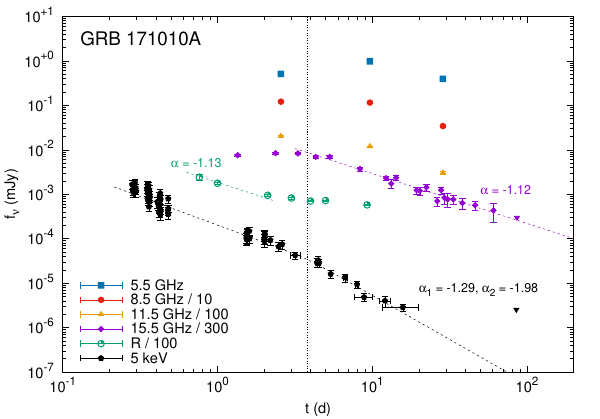}
\includegraphics[width=0.95\columnwidth]{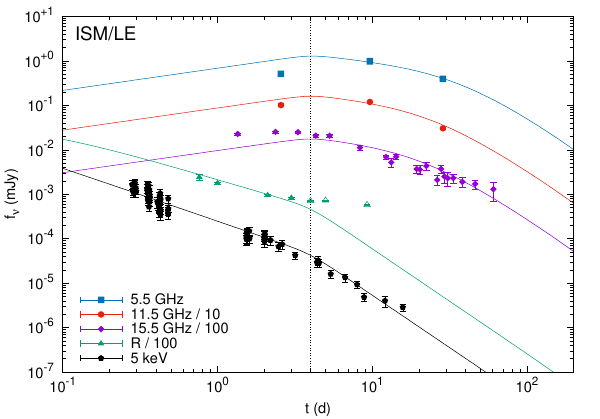}
\includegraphics[width=0.95\columnwidth]{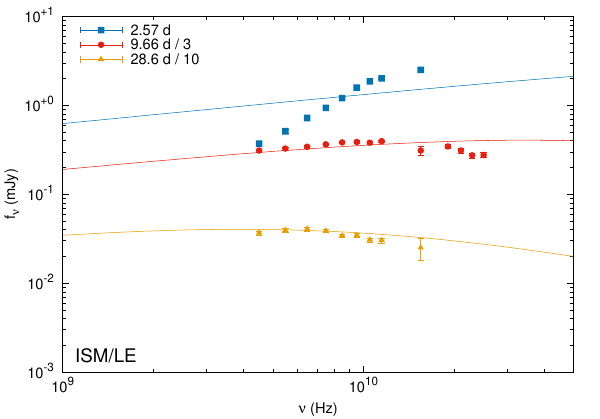}
\caption{Our single and broken power law fits to the light curves of GRB 171010A (upper panel), and our best-fitting analytical model compared to the light curves (middle panel) and radio SEDs (lower panel). Ten additional optical and radio frequencies used in the fitting are omitted from the light curve figure for clarity.}
\label{fig:171010a}
\end{figure}

An X-ray break was seen for this burst, but the radio behavior (Fig. \ref{fig:171010a}) was described as unusual by \citet{bright19}. They attempted to explain the light curve evolution using a steep CBM density profile, but nonetheless found that the evolution of break frequencies in the spectrum seems too slow for the standard model to explain. The X-ray light curve is consistent with $p\approx 2.4$ with an edge-effect break, while the optical decline and the spectral index of $\beta=-0.87$ between optical and X-ray together imply that $R < \nu_c < \mathrm{5~keV}$ in ISM. The radio decline slope of $\alpha = -1.12\pm0.05$ is consistent with $p\approx2.5$ and pre-break. However, the peak of the radio light curve is roughly simultaneous with the X-ray break, and thus GRB 171010A seems to exhibit another delayed radio break; we can place a limit of $t_{j, \mathrm{radio}} > 13 \times t_{j, \mathrm{X}}$ in this case based on the last radio detection. This requires that $\nu_m < 15.5$ GHz at $t_{j, \mathrm{X}}$, while the first SED suggests $15.5~\mathrm{GHz} < \nu_m$ and $\nu_a \sim10$~GHz. As \citet{bright19} point out, if the peak frequency is $\nu_m$, its evolution is also slower than expected. No post-jet-break scenario allowed by the X-ray and optical is consistent with the radio decline.

Our own best-fit model agrees with the findings of \citet{bright19}: individual SEDs can be reproduced by it, but not all epochs simultaneously. Specifically, our best-fit model disagrees with the earliest SED ($\sim2.5$~d). As for the light curve, instead of a single power law, our best fit is a smooth transition from $\nu_a < 15.5~\mathrm{GHz} < \nu_m$ to $\nu_m < \nu_a < 15.5~\mathrm{GHz}$. This prediction, however, deviates from the observed pre-$t_j$ 15.5~GHz light curve. At other frequencies the light curve is sparse and such comparisons are more difficult. We consider GRB 171010A inconsistent with the standard model.

\subsection{GRB 990510}

\begin{figure}
\centering
\includegraphics[width=0.95\columnwidth]{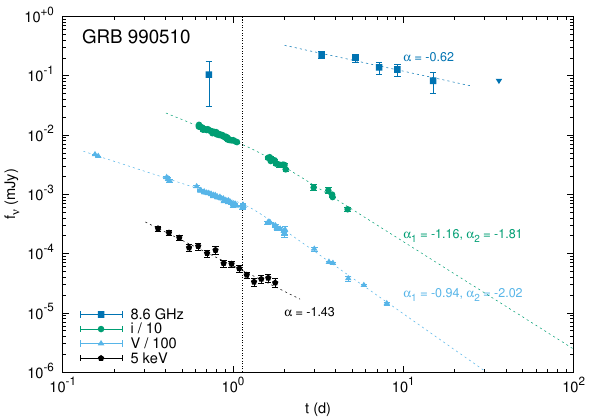} 
\includegraphics[width=0.95\columnwidth]{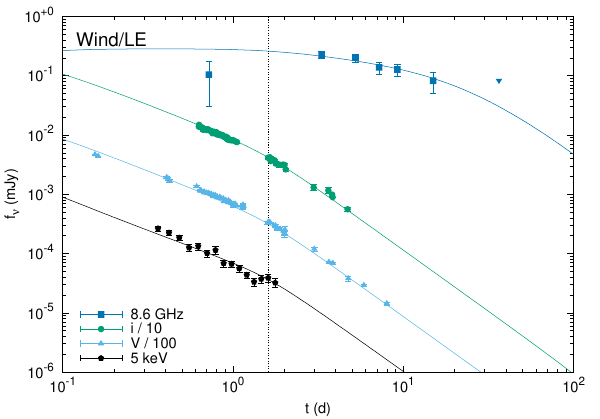}
\caption{Our single and broken power law fits (upper panel) and best-fitting analytical model (lower panel) compared to the light curves of GRB 990510. Four additional optical and radio frequencies used in the fitting are omitted from the light curve figure for clarity.}
\label{fig:990510} 
\end{figure}

For GRB 990510, only the optical shows evidence of a jet break (Fig. \ref{fig:990510}), but as the X-ray follow-up was much shorter, a break cannot be excluded in the X-ray. Both the optical and X-ray pre-break light curves are consistent with $p\approx2.4$ and $\nu_m < \nu_c$ in an ISM-type CBM, but in this case the optical post-break decline may require a jet break mechanism that involves a combination of the edge effect and logarithmic lateral expansion, as seen in simulations by \citet{veerten12}. However, the post-optical-break radio decline can be fitted with $\alpha_{\mathrm{radio}} = -0.62 \pm 0.10$, which is not compatible with any pre- or post-break slopes allowed by the higher frequencies. Our best-fit analytical model is able to reproduce the radio and optical light curves with a smooth transition from $\nu_a < 8.6~\mathrm{GHz} < \nu_m$ to $\nu_m < \nu_a < 8.6~\mathrm{GHz}$, but in this case a $t^{-p}$ decline is predicted soon after the last radio observations. We conclude that GRB 990510 is consistent with the standard model if a $t^{-p}$ decline began soon after the radio observations ended. A similar quality fit was obtained by \citet{pk01}, but as they use different equations for break frequencies and flux, they arrive at different parameters.

\subsection{GRB 991208}

\begin{figure}
\centering
\includegraphics[width=0.95\columnwidth]{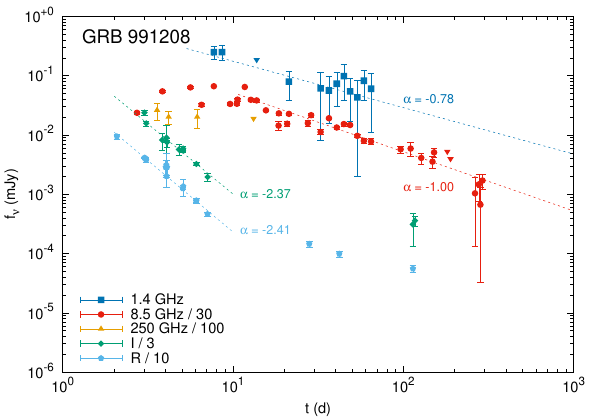} 
\includegraphics[width=0.95\columnwidth]{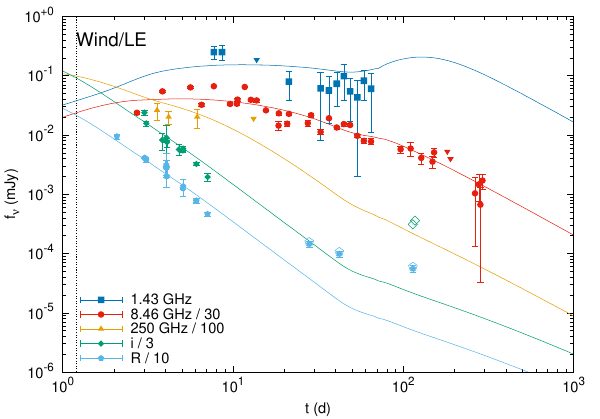}
\includegraphics[width=0.95\columnwidth]{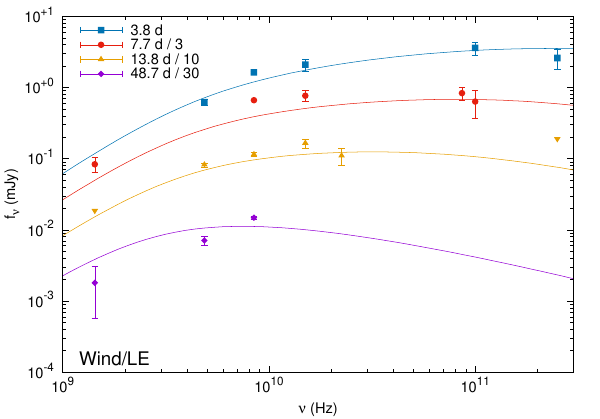}
\caption{Our single and broken power law fits (upper panel) to the light curves of GRB 991208, and our best-fitting analytical model compared to the light curves (middle panel) and radio SEDs (lower panel). Nine additional optical and radio frequencies used in the fitting are omitted from the light curve figure for clarity.}
\label{fig:991208} 
\end{figure}

No X-rays were detected for this GRB, but we can use the optical light curve (Fig. \ref{fig:991208}) to infer that $p\approx2.4$ -- assuming lateral expansion -- and that the jet break happened some time before the optical observations ($\lesssim2$~d). Without lateral expansion, assuming $p<3$, the optical slope implies $\nu_a, \nu_c, \nu_m < \nu$ in ISM or $\nu_a, \nu_m < \nu < \nu_c$ in wind. The 8.5 GHz light curve is consistent with post-jet-break in the $\nu_a < \nu_c, \mathrm{8.5~GHz} < \nu_m$ scenario, but also with a pre-jet-break slope with $p=2.4$, $\nu_a, \nu_m < \mathrm{8.5~GHz} < \nu_c$ in ISM. The spectrum, however, implies that $\mathrm{max}(\nu_m,\nu_a) >$~10 GHz until at least $\sim50$~d. The slope continues until $\geq 300$~d, i.e. in the pre-jet-break case we can set a limit of $t_{j,\mathrm{radio}} \gtrsim140 t_{j,R}$ using the last radio detection. In a wind-type CBM, the radio slope is consistent with post-jet-break expectations, $t^{-1}$, if the fast-cooling spectrum (i.e. $\nu_m > \nu_c$) persists until late times.

Our best-fit model (with wind-type CBM and lateral expansion) can reproduce the shapes of the SEDs and the light curves reasonably well -- albeit with some deviation at 1.43 GHz -- with a smooth transition from $\nu_a < 8.6~\mathrm{GHz} < \nu_m$ to $\nu_m < \nu_a < 8.6~\mathrm{GHz}$, followed by the onset of the non-relativistic phase at $\sim50$~d. It does slightly over-predict the optical points at $\sim7$~d, though. Thus we conclude that GRB 991208 is tentatively consistent with the standard model. The preferred $\epsilon_e = 0.83_{-0.10}^{+0.08}$ is rather high, but setting an upper limit of $\epsilon_e < 0.7$ results in a fit of only slightly worse quality. It is worth noting, however, that this burst has no pre-jet-break data in any band, nor any X-ray data.

\subsection{GRB 000301C}

GRB 000301C was also not detected in X-rays, but a jet break was observed in the optical bands; the optical slope of $\sim t^{-2.6}$ (in the $R$ band, with by far the longest follow-up among the optical bands) is too steep for an edge effect break, and thus implies $p\approx2.6$. The $K$- and $B$-band post-break slopes are not as steep, but the follow-up is shorter, which may affect the result. \citet{berger00} note fluctuations in the optical light curves that can explain the different fits to the different frequencies; these may be due to density changes in the CBM. Indeed, they (and our best-fit model below) suggest that the jet break occurred around the last $K$-band observation, and the "break" in $K$ is merely the peak of a bump in the light curve. The perturbed optical light curve may also explain the large difference between pre-and post-jet-break slopes: in $B$ and $R$, the pre-jet-break slope is $\sim-0.6$, seemingly incompatible with $p=2.6$, and the $K$-band decline is even shallower. 

The radio decay seems much steeper at 4.9 than 8.6 GHz, but this is plausibly due to the smaller number of data points, none of them between 10 and 40 d, with much larger errors. The better-sampled 8.6 GHz decay can be fitted with a single power law from $\sim10$ to $\gtrsim100$~d. This power law ($t^{-0.74\pm0.07}$) is not consistent with post-jet-break expectations with lateral expansion, but within $1.1\sigma$ of the pre-break $t^{-2/3}$ in a wind CBM and $8.6~\mathrm{GHz} < \nu_c < \nu_m$ until $\gtrsim20 t_{j,R}$.

Our best-fit model does a good job of reproducing most of the radio and optical light curve (allowing for some optical deviation because of the aforementioned fluctuation) with a smooth post-jet-break $\nu_m$ passage. However, it has difficulty with the pre-$t_j$ 250-GHz data, which it consistently under-predicts, while at 4.2 d, it places $\nu_a$ at a much lower value than the observations indicate. \citet{berger00} were able to better reproduce the radio SED, but the break frequencies in their model were free parameters instead of being determined by physical parameters. Thus, despite the lack of X-ray data to constrain the GRB, we consider GRB 000301C problematic for the standard model. \citet{p05} used a structured jet model to describe GRB 000301C, but their fit is of similar quality as ours in the optical and worse in the radio.

\subsection{GRB 000926}

\begin{figure}
\centering
\includegraphics[width=0.95\columnwidth]{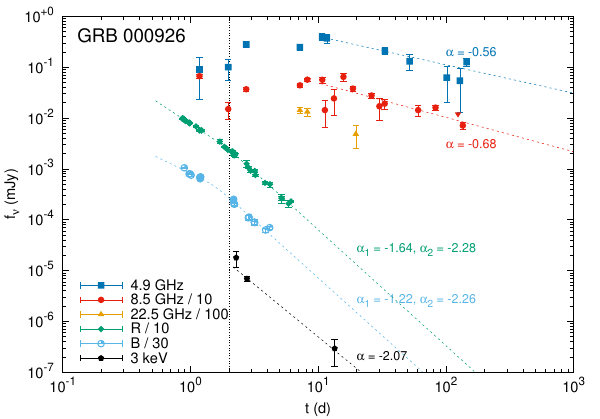} 
\includegraphics[width=0.95\columnwidth]{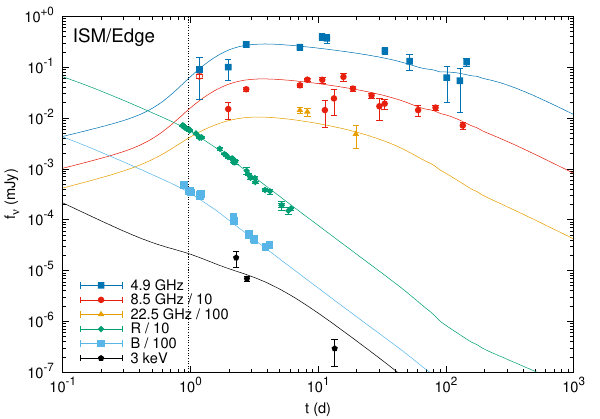}
\includegraphics[width=0.95\columnwidth]{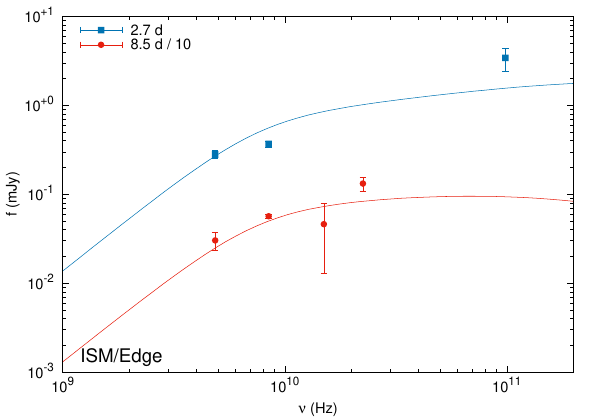}
\caption{Our single and broken power law fits (upper panel) to the light curves of GRB 000926, and our best-fitting analytical model compared to the light curves (middle panel) and radio SEDs (lower panel). Seven additional frequencies used in the fitting are omitted from the light curve figure for clarity.}
\label{fig:000926} 
\end{figure}

In the case of GRB 000926, the jet break seems to have been captured in the optical light curve, which is consistent with an edge effect jet break in ISM- or wind-type CBM at $\sim1.4\sigma$ and $\sim1.8\sigma$ respectively ($\Delta \alpha = -0.64\pm0.08$). The ISM scenario is disfavored by the pre-break optical decay. The best-sampled band, $R$, is consistent with $p\approx2.5$ ($\nu_m < R < \nu_c$) or $p\approx2.8$ ($\nu_c < \nu_m < R$) in wind, but requires $p>3$ in ISM. The radio light curve decline is consistent with being post-jet-break in wind with edge effect and $\nu_a < \nu < \nu_m < \nu_c$. The decline is also close to the expected \emph{pre}-jet-break slope in a wind CBM in the fast-cooling scenario, which is plausible based on the optical and X-ray data. The radio light curve is rising or flat (the apparent scintillation at this time makes it difficult to tell) until $\sim 5 t_{j,R}$, which is consistent with the above if the slope during the rise is $t^{1/2}$ and thus $\nu < \nu_a$.

Our best-fit model generally reproduces the shape of the optical and radio light curves. \citet{harrison01} and \citet{pk02} argued that the X-ray requires a contribution from IC emission, while \citet{p05} was unable to explain the late-time radio behavior with a structured jet model. Our best fit in terms of $\chi^2$ does not require IC emission, but instead imposes $\epsilon_B \approx 1$ -- this requirement for models without IC effects was also noted by \citet{harrison01}. By requiring a more typical value, $\epsilon_B < 0.1$, our model also needs IC emission; in this light we show this fit in Figure \ref{fig:000926} and in Table \ref{tab:models}. The shape of the radio and optical light curves is reproduced well, but there is a $\sim3.5\sigma$ deviation at the last X-ray epoch. The X-ray light curve is sparse. Overall, we find GRB 000926 tentatively consistent with the standard model.
    
\subsection{GRB 100418A}
   
\begin{figure}
\centering
\includegraphics[width=0.95\columnwidth]{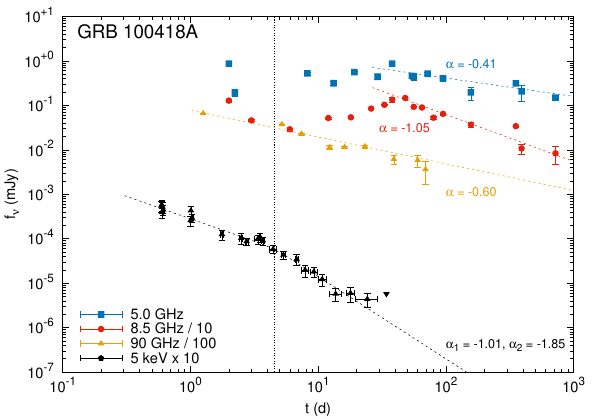}
\includegraphics[width=0.95\columnwidth]{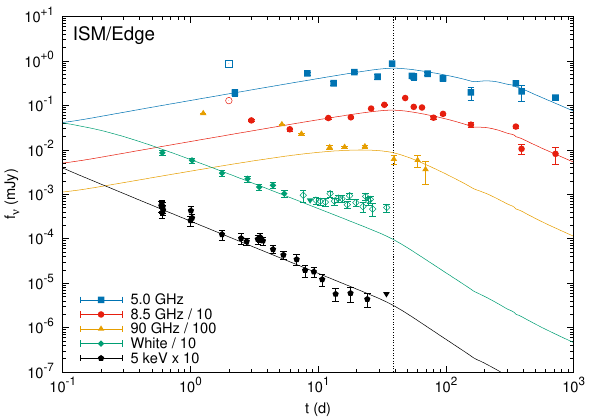}
\includegraphics[width=0.95\columnwidth]{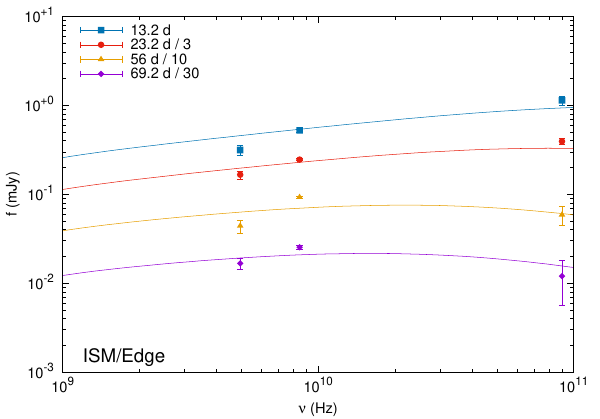}
\caption{Our single and broken power law fits (upper panel) to the light curves of GRB 100418A, and our best-fitting analytical model compared to the light curves (middle panel) and radio SEDs (lower panel). Eight additional optical and radio frequencies used in the fitting are omitted from the light curve figure for clarity.}
\label{fig:100418a}
\end{figure}

Based on the X-ray light curve (Fig. \ref{fig:100418a}), we infer $p\approx2.0$ and $\nu_c < \nu$, or alternatively $p\approx2.3$ and $\nu_m < \nu < \nu_c$; in the latter case the CBM is restricted to ISM-like. The value of $\beta = -1.00\pm0.05$, reported by \citet[][]{dup18} to fit the spectrum from optical to X-ray at $\sim2$ d, rules out the former unless $\nu_c$ is also below the optical bands. The radio light curve at 5 and 8.5 GHz peaks around 50 d ($10\times t_{j,X}$) and at each frequency settles onto a power law decay, though the power-law index varies substantially with frequency. At 90 GHz, the maximum of the light curve takes place at $t \lesssim 1$ d, and the slope thereafter is consistent with pre-jet-break and $p \approx 1.8$, meaning $t_{j,90~\mathrm{GHz}} \gtrsim 15t_{j,X}$ if the slope is indeed pre-break. With such a low $p$ we also need to use Eqs. (4)--(7) in \citet{daicheng01} for $p<2$. We find that the 90 GHz and X-ray light curves are both consistent with $p\approx1.5$ and an ISM-type CBM, with $\nu_c < 5~\mathrm{keV}$, and $\nu_m < 90~\mathrm{GHz} < \nu_c$, but the other radio frequencies are not. No other $p<2$ scenario is consistent with both X-ray and millimeter frequencies.

\citet{moin13} suggested that the late peak epoch seems to require a model of late energy injection that revitalizes the forward shock. With a re-energized shock one would still expect the evolution beyond the late-time peak to resemble the post-peak evolution of the 'standard' case; but a post-jet-break decline is not seen at any point in the radio. It is, however, possible that there is a second peak in the 90 GHz light curve around 25 d, in which case the slope before its onset could in principle be steeper; but this is not seen convincingly in the data. The peaks in the radio light curves seem to roughly correspond to $\nu_m$ passage based on the SED evolution; thus the late injection model remains plausible.

\citet{laskar15} attempted to model this burst using an energy injection episode at $\lesssim0.5$~d; they considered strong scintillation a possible explanation for the shape of the radio light curve between 2 and 20 days, and ignored these data points in their fitting process. By doing this, they were able to fit the light curve. They did not, however, have access to 4.8 GHz data in the same time period from \citet{dup18}, which behave roughly the same way as the 8.5 GHz data. We attempt a fit as well (ignoring the very bright first points of the 5.0 and 8.5 GHz light curves instead, along with the pre-injection points at $<0.5$ d), and find that most of the features of this GRB can be roughly reproduced, but the 90 GHz light curve at $<10$~d is clearly severely under-predicted until 10 d. We find GRB 100418A inconsistent with the model, unless the excess is produced by a reverse shock that is hidden by forward shock emission at lower frequencies. The 90 GHz decline is slow for an RS, but may be transitioning from $\nu < \nu_m^{\mathrm{RS}}$ to $\nu_m^{\mathrm{RS}} < \nu$. \citet{laskar15} disfavored a reverse shock in this GRB, but did not explore it in detail. We note that, as in \citet{laskar15} and \citet{marshall11}, our fit places the jet break at a much later time than the broken power law indicates.

\subsection{GRB 111215A}
   
A single pre-jet-break power law consistent with $p\approx2.4$ and $\nu_c < \nu$ fits the X-ray data of this dark GRB until $t \sim 20$ d. Around this time, the 93 GHz light curve exhibits an unambiguous break if the upper limit at 41 d is considered. An X-ray break is not seen, but may have occurred around 20~d as well. The other radio light curves decay as single power laws consistent with ISM and $\nu_a < 4.9~\mathrm{GHz} < \nu_c < 19.1~\mathrm{GHz} < \nu_m < 93~\mathrm{GHz}$.

Based on the spectrum, $\mathrm{max(\nu_m,\nu_a)}$ seems to be located above 5 GHz until $\gtrsim26$ d, perhaps as late as 45 d; while the light curve at 4.9 GHz already seems to decline at this point. A decline is, however, allowed if $\nu_a < \nu_m$ and the X-ray light curve breaks around the same time as the 93 GHz one. The 93 GHz break is roughly simultaneous with the peak at $\lesssim20$~GHz and the end of X-ray observations, which also points toward an achromatic transition.

\citet{zauderer13} used the standard model with a jet break around the last X-ray observations; however, this does not provide a good fit to the late-time radio points. \citet{vdh15} were able to fit the afterglow, placing a lower limit of $>31$~d for the jet break. They, however, did not consider the effects of IC cooling (or emission), which is important with the ratio of $\epsilon_e/\epsilon_B \approx 2400$ of their best fit. Our model, which does include IC effects, can account for the X-ray and some of the radio light curve; however, our best ISM fit does not reproduce the 93 GHz light curve shape or the shape of the radio SED, while our wind fits, slightly better in terms of $\chi^2$, also cannot meet the upper limits at 1.4~d and $\leq7$ GHz. Thus we consider GRB 111215A inconsistent with our model.
    
\subsection{GRB 140311A}
   
\begin{figure}
\centering
\includegraphics[width=0.95\columnwidth]{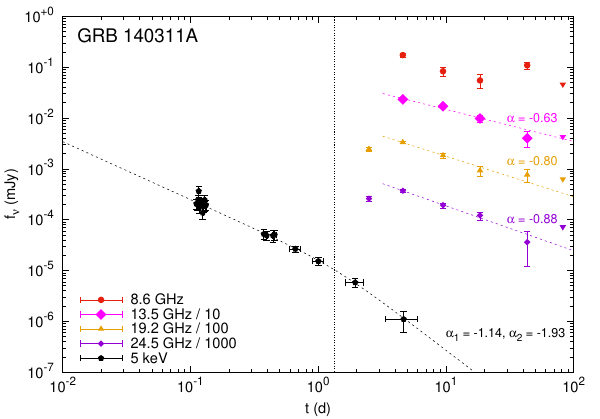}
\includegraphics[width=0.95\columnwidth]{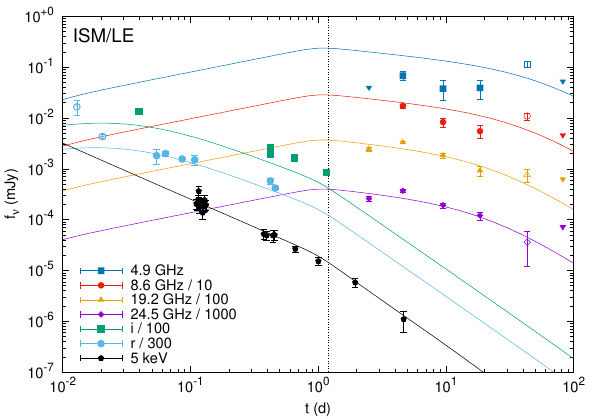}
\includegraphics[width=0.95\columnwidth]{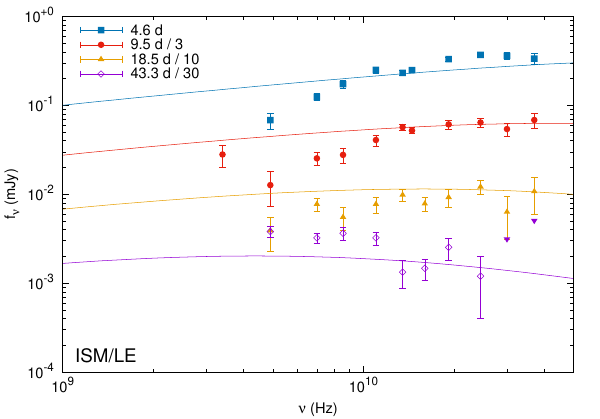}
\caption{Our single and broken power law fits to the light curves of GRB 140311A (upper panel), and our best-fitting analytical model compared to the light curves (middle panel) and radio SEDs (lower panel). Nine additional optical and radio frequencies used in the fitting are omitted from the light curve figure for clarity.}
\label{fig:140311a}
\end{figure}

The presence of an X-ray break is ambiguous ($P_F = 0.85)$, but the X-ray light curve (Fig. \ref{fig:140311a}) is consistent with $p\approx2.2$ and $\nu_c < \nu$; the large error on the post-break slope also makes it consistent with either lateral expansion or an edge effect. The radio light curves are mostly single power laws consistent with pre-break, $p\sim2.1$ and $\nu_m < 8.6~\mathrm{GHz}$; in this case we can set a limit of $t_{j,\mathrm{radio}} \gtrsim 3t_{j,X}$. However, the light curve at $\leq 8.6$ GHz peaks again at $\sim40$ d, which is at odds with the expected behavior and with other frequencies. 
    
The SED indicates $\nu_m$ is located at $\gtrsim25$~GHz when the 24.5 GHz light curve peaks ($\sim5$~d). However, the 8.6 and 13.5 GHz light curves also decline from this point on, while $\nu_m$ does not pass through 13.5 GHz until $\gtrsim20$ d. A decline below $\nu_m$ is allowed by standard theory after the jet break, but in this case the slope of the light curve is steeper at all frequencies than the expected $-1/3$. 

Our best-fit model is unable to reproduce the behavior of the radio SED, even when the late-time bump at 43 d is ignored. We also ignore the sharply declining optical feature at $<0.03$~d. The optical, X-ray and $\gtrsim10$~GHz light curves can be fitted reasonably well, but the model fails at the lowest observed frequencies ($\lesssim 9$~GHz). \citet{laskar18a} were able to fit the radio SED better (namely, their $\nu_a$ is located around 10 GHz), but cannot simultaneously fit the X-ray decline well. We can roughly reproduce their fit using their parameters. They consider the short-lived bump at $\sim40$~d as possibly an early transition to a non-relativistic phase caused by a density enhancement. In summary, we consider GRB 140311A inconsistent with the standard model.
    
\subsection{GRB 140903A}

\begin{figure}
\centering
\includegraphics[width=0.95\columnwidth]{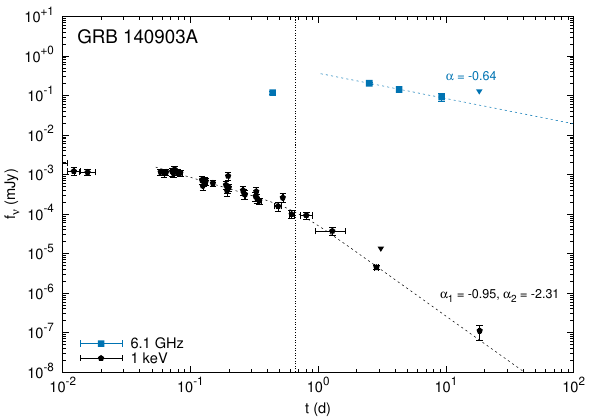} 
\includegraphics[width=0.95\columnwidth]{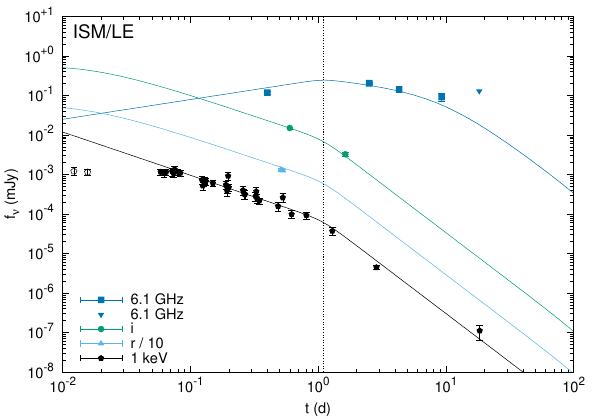}
\caption{Our single and broken power law fits (upper panel) and best-fitting analytical model (lower panel) compared to the light curves of GRB 140903A.}
\label{fig:140903a}
\end{figure}

This short GRB is exceptional in its class in that it has an observed X-ray jet break \citep{troja16}. The X-ray light curve (Fig. \ref{fig:140903a}) is consistent with $p\approx2.3$ and $\nu_m < \nu < \nu_c$ in an ISM-like CBM. However, the radio light curve at 6.1 GHz is inconsistent with this and would, instead, require a wind-type CBM (unlikely for short GRBs) and a fast-cooling spectrum ($\nu_m > \nu_c$) to be compatible with the standard model. We do note that the radio data are rather scarce for this burst. 

Our best-fit model reproduces all light curves reasonably well, but the radio light curve does steepen around the last observed epoch. \citet{troja16} consider this burst as slightly off-axis. Even on-axis, their reported parameters, albeit not the best fit in terms of $\chi^2$, reproduce the data fairly well by eye. Thus we conclude that GRB 140903A remains consistent with the standard model.
    
\subsection{GRB 161219B}
   
\begin{figure}
\centering
\includegraphics[width=0.95\columnwidth]{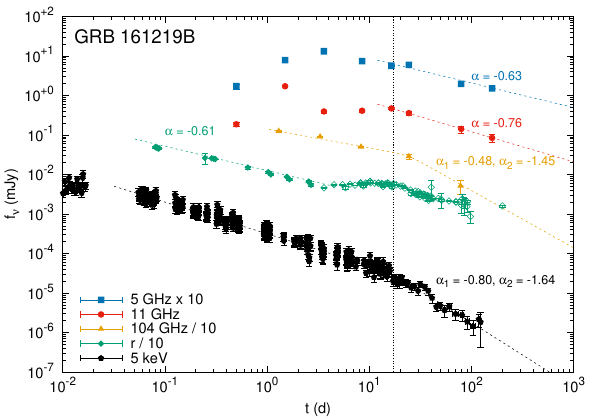}
\includegraphics[width=0.95\columnwidth]{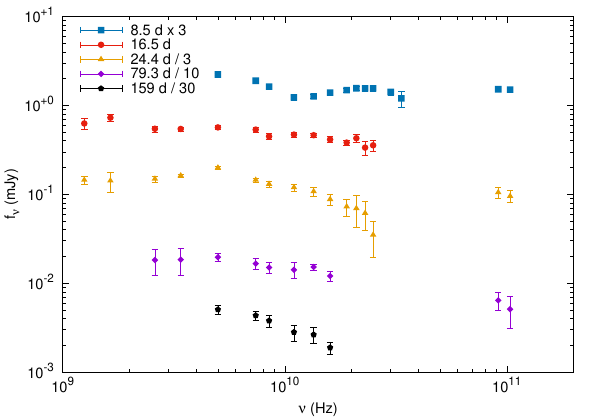}
\caption{Our single and broken power law fits to the light curves of GRB 161219B (upper panel), and the evolution of the radio SED of this burst (lower panel).}
\label{fig:161219b}
\end{figure}

The X-ray decline can be fitted with a broken power law, with a post-break slope of $-1.64\pm0.11$ (Fig. \ref{fig:161219b}). From the pre-break slope of $-0.80\pm0.01$ one obtains $p\approx1.7$, $\nu_c < \nu$; in which case an ISM-type CBM with either an edge-effect or lateral expansion break is a good match. The initial decline at 104 GHz is also consistent with this using $p\approx1.7$ and $\nu_m < \mathrm{104~GHz} < \nu_c$. A value of $p$ close to this ($p\approx1.8$) also matches the $r$-band pre-jet-break decline of $-0.61\pm0.01$ (we cannot fit the optical post-jet-break afterglow as it is dominated by SN2016jca and a late-time host galaxy contribution). However, the standard equations should not apply when $p<2$. The \citet{daicheng01} equations, which should apply here, are not consistent with the early X-ray and optical slopes with any $p$ between 1 and 2. Furthermore, the radio light curve \citep[after the proposed reverse shock no longer dominates;][]{laskar18c} at the frequencies with the longest follow-up settles onto a power law inconsistent with this scenario. If $\nu_X > \nu_c$, on the other hand, the early X-ray slope is consistent with $p\approx2.1$, but the optical light curve is not. The 104 GHz curve seems to break at or after $\sim20$ d; there are not enough post-break epochs for a proper broken power law fit, but simply using the last two points, one obtains a post-break slope of $\alpha_{2,\mathrm{104 GHz}} = -1.47\pm0.31$, consistent with the X-ray slope. A steeper slope is also possible if the break occurs later, though. This break is not seen at lower frequencies. 

Our analytical model, being based on \citet{granotsari02}, is naturally unable to handle values of $p<2$ that seem to be required by the optical and X-ray light curves. The model invoked by \citet{laskar18c} requires a refreshed reverse shock; however, even this model somewhat under-predicts the flux at late times, after their proposed jet break time. Additionally, \citet{laskar19} pointed out that the millimeter light curve of their model deviates somewhat from the observed light curve of this burst and of GRB 181201A. The spectrum does seem to require multiple peaks at least until 24.4~d, but the late-time spectra at $\geq79.3$~d are consistent with $\nu_m$ being located at $\sim5$~GHz. All in all, the radio emission is inconsistent with the standard model, and only (roughly) consistent with the proposed refreshed reverse shock model shown in \citet{laskar18c} if the $t^{-p}$ decline started soon after the last radio observations. \citet{laskar18c} also briefly bring up the possibility of either a host galaxy contribution to the late-time radio flux -- which, however, is already included in said model -- or an early transition to a non-relativistic phase, similarly to \citet{laskar18a}.
 
\section{Discussion} \label{sec:disco}

\subsection{Anomalous GRBs and jet break effects}

As the jet break is a geometric effect, in standard GRB jet theory we should see a post-jet-break slope associated with it at all frequencies -- even when the radio emission source is a population of thermal electrons behind the shock wave as suggested by e.g. \citet[][]{eichler05,ressler17,warren18}. Several different power laws can be expected depending on the CBM profile, the presence of lateral expansion and the ordering of the break frequencies in the spectra \citep{rhoads99,mrees99,granotsari02}, as summarized in our Table \ref{tab:stdmodel}. It is also possible to have a scenario where some lateral expansion is present, but the break is dominated by the edge effect \citep[this was demonstrated in simulations by][]{veerten12}. However, based on a large fraction of events of our sample, the situation may be more complicated in the radio.
 
\begin{table*}
\centering
\begin{minipage}{\linewidth}
    \renewcommand{\arraystretch}{0.9}
    \begin{tabular}{lll}
    \centering
       GRB & Radio behavior & Model fit \\
        \hline
       Group 1 \\
       \hline
        050820A & SPL consistent with pre-break until $\gtrsim 4t_{j,X}$\footnote{These limits are based on the last radio detection of each target.} & Consistent if very low density and opening angle\\
        051022 & SPL consistent with $\nu_m > 4.9$ GHz & Good\\
        070125 & SPL inconsistent with theory; late peak at low & Under-predicted 22.5 and 95 GHz light\\
         & frequencies & curves \\
        090313 & SPL consistent with $\nu_m > 16$ GHz, but too long & Over-predicted radio light curves; power\\
         &  & law not reproduced\\
        110709B & SPL consistent with pre-break until $> 60t_{j,X}$ & Good\\
        120326A & BPL & Under-predicted early X-ray light curve \\
        130907A & SPL consistent with pre-break until $> 170t_{j,X}$ & Under-predicted late radio light curves\\
        141121A & Complex; multiple peaks & ... \\
        151027A & Hint of a break at $\sim100$~d & Consistent if $p \approx 3.0$\\
        160509A & SPL consistent with pre-break until $> 20t_{j,X}$ & Good, but requires steepening after observations\\
        160625B & SPL consistent with pre-break until $> 10t_{j,X}$ & Over-predicted late 6.1~GHz light curves;\\
         &  & power law not reproduced\\
        171010A & SPL consistent with pre-break until $> 13t_{j,X}$; $\nu_m$ & Early radio spectrum not reproduced \\
         & evolution too slow & \\
         \hline
       Group 2 \\
       \hline
        990510 & SPL inconsistent with theory & Good, but requires steepening after observations\\
        991208 & SPL consistent with pre-break until $\gtrsim 140t_{j,R}$ & Fair \\
        000301C & SPL consistent with pre-break until $\gtrsim 20t_{j,R}$ & Under-predicted mm light curve; over- \\
         &  & predicted early 4.9 and 8.5 GHz fluxes\\
        000926 & SPL consistent with post-break & Fair, but IC emission needed in X-rays \\
        100418A & SPL inconsistent with theory; late peak at low & Under-predicted early mm light curve \\
         & frequencies & \\
        111215A & BPL (93 GHz) or SPL consistent with post-break & Over-predicted early $<10$ GHz light curves\\
        140311A & SPL consistent with pre-break until late-time re-brightening; & Over-predicted $<10$ GHz light curves\\
         & SED inconsistent with light curve & \\
        140903A & SPL only consistent with $\nu_m > \nu_c$ until late times & Fair, but requires steepening after observations\\
        161219B & BPL ($\sim100$ GHz) or SPL inconsistent with theory & ... \\
        \hline
    \end{tabular}
    \caption{Summary of our main findings. For each GRB, we briefly describe the radio behavior based on our single and broken power law fits, and whether it is consistent with the X-ray behavior in some scenario of the standard jet model. We also describe whether the best-fit analytical model is a reasonable match with the observations.}
    \label{tab:scenarios}
\end{minipage}
\end{table*}

In Table \ref{tab:scenarios}, we summarize how the radio light curve and SED evolve and whether they are consistent with the higher frequencies; we also describe how our analytical model fits the observations. For most of the GRBs in our sample, at late times we find a power-law decline in the radio whose slope is inconsistent with that expected based on the post-break X-ray or optical afterglow. In a few cases in our sample, we do see a post-jet-break radio slope predicted by the standard model and consistent with the same CBM type, spectral regime and $p$ as the optical and X-ray; these are GRBs 090313 and 120326A in Group 1 and GRBs 000926 and 111215A in Group 2. More commonly, we see a decline consistent with \emph{pre}-jet-break predictions until $>10$ or even $>100$ times later than the break seen in the X-rays or optical (namely, in GRBs 050820A, 110709B, 130907A, 160509A, 160625B and 171010A in Group 1 and GRBs 991208 and 000301C in Group 2), or inconsistent with all predicted slopes. As \citet{veerten11} showed, break times may differ on different sides of $\nu_a$ and the break may broaden below $\nu_m$, but in these simulations this effect was by a factor of a few, and in any case cannot explain the continuing pre-jet-break slope when the radio behavior is consistent with $\nu_m < \nu$. In six of the afterglows in our sample, with radio data both before and after the jet break, we see a simultaneous steepening of the radio light curve or at least cannot exclude it, indicating that the jet break does affect the radio (these are GRBs 990510, 050820A, 140903A, 160625B, 161219B and 171010A, while only GRBs 051022 and GRB 100418A do not show this), but the following decline is still generally inconsistent with predicted (asymptotic) power-law slopes. Furthermore, the radio light curves of eight GRBs (070125, 090313, 110709B, 130907A, 991208, 000301C, 000926 and 140311A) clearly continue rising after the X-ray jet break.

In the sample, we see one case where a $\sim t^{-2}$ decline is observed at $\sim10$~GHz (GRB 120326A) and two others where it is seen at millimeter frequencies (GRBs 111215A and 161219B). The radio decline of GRB 120326A before the $\sim t^{-2}$ decline is seemingly inconsistent with expectations if the steepening corresponds to a post-jet-break passage of $\nu_m$; however, a smooth transition is able to reproduce the shape of the light curve (although the early X-ray fluxes are simultaneously under-predicted). The light curve scenarios described above and in K20, with a seemingly single power law inconsistent with any asymptotic power law in the standard model, can, therefore, at least in some cases simply be the result of an insufficiently long follow-up. In almost all cases, the constraints we have on the evolution of $\nu_m$, based on the timing of its observed passage (including the aforementioned steepening in a millimeter light curve), are not strict enough to say a break onto a $\sim t^{-2}$ decline should have happened. This is not always the case, however: in GRBs 090313 and 130907A our code is unable to reconcile the lack of a $\sim t^{-2}$ decline with other frequencies. In four cases (GRBs 990510, 110709B, 140903A and 160509A), the model light curve is consistent with the data but steepens around the last observed radio point. 

A caveat exists for the GRBs where the standard model cannot be excluded. Whether the single or broken power law of the radio decline is consistent with theoretical predictions is not an indicator of whether our code can find a reasonable fit for it. For example, GRB 111215A is found inconsistent with the standard model by our code despite the light curve slopes, while GRB 160509A is consistent with it. Smooth transitions between power law segments can account for some of the inconsistencies, while the GRBs with seemingly consistent slopes may be ruled out through other constraints. As per Table \ref{tab:scenarios}, no systematic difference is seen in the overall radio behavior of the GRBs consistent and inconsistent with the model.

As our Group 1 constitutes a representative sample, we can use it to study the fraction of the general (radio-loud) GRB population that is consistent with the standard model. Out of the 12 objects in Group 1, we find three (GRBs 051022, 110709B and 160509A) to be unambiguously consistent with the model, while two (GRBs 050820A and 151027A) are consistent with it if one allows unusual parameter ranges. Counting these two, we thus arrive at five events consistent with the model, out of 12; this means a fraction\footnote{Here, we have used \citet{gehrels86} to determine the uncertainty of the fraction.} of $0.42_{-0.17}^{+0.19}$ for the well-fit afterglows out of all radio-loud GRBs. This number does not change when one checks the best fits for Klein-Nishina effects \citep{nakar09}; see Appendix \ref{sec:app2}. This implies that in a significant fraction of GRBs, the physical parameters (e.g. CBM density and jet energy) determined through comparison to the standard model may be incorrect, in turn possibly affecting our understanding of the mechanism of the central engine. In the non-representative Group 2, we are able to fit four afterglows out of nine -- these include two events in our sample that were included in \citet{pk04}, where their radio decline was considered anomalously shallow (GRBs 991208 and 000926). In both of these cases, a transition to the non-relativistic phase can explain the behavior. A non-relativistic transition was also suggested by \citet{frail04} to explain the late-time flattening they observed in some GRBs. While our modeling results indicate that flatness or shallowness of the radio decline is not \textit{necessarily} a problem, it also suggests that non-relativistic transitions can not explain cases such as GRBs 160625B or 171010A.

\subsection{Secondary emission component}
\label{sec:2comp}

One explanation for the long-lasting single power-law decline of the radio emission might be the presence of a wider jet surrounding a more energetic narrow core, which dominates in the radio \citep[similar to that suggested by][]{berger03}. The greater width of this jet would produce a later break time than that expected from the narrower 'main' jet. Another separate source of radio-dominated emission could be the injection of energy into the reverse shock in the form of slower ejecta catching up to it, suggested by \citet{sm2000} and \citet{pk04}\footnote{\citet{laskar18c} also suggest a reverse shock with energy injection for GRB 161219B, but not to explain the late-time radio light curve.}. However, if one of these scenarios applies to a GRB and the radio emission is dominated by a second source at a certain epoch, the radio emission associated with the narrow jet must then be significantly weaker than the observed flux at that epoch.

\begin{figure}
\centering
\includegraphics[width=\columnwidth]{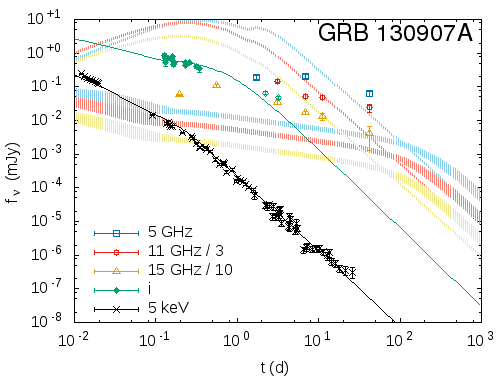} 
\includegraphics[width=\columnwidth]{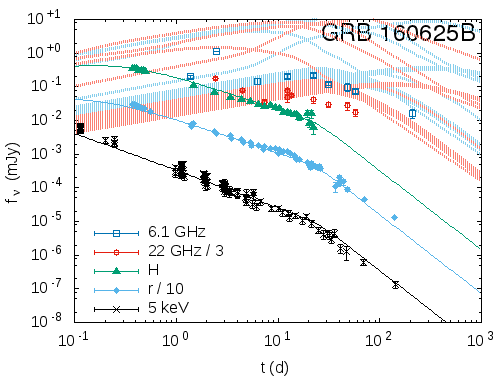} 
\caption{The best-fitting X-ray and optical light curves and the range of variation in radio light curves for 500 walkers in GRBs 130907A and 160625B, when fitting without radio data. The radio fluxes of GRB 130907A are clearly under-predicted apart from one outlying walker, while those of GRB 160625B are close to the observed fluxes. Thus GRB 130907A can have a dominant contribution from a secondary component without suppressing the flux from the narrow jet.}
\label{fig:2comp_test}
\end{figure}

To investigate the plausibility of producing faint radio emission from the narrow jet within the confines of the standard model, we have re-run our model-fitting code for the bursts in our sample with relatively good X-ray and optical coverage (GRBs 990510, 050820A, 070125, 090313, 120326A, 130907A, 140311A and 160625B)\footnote{GRB 151027A was not included as, despite being tentatively consistent with the standard model, it is possible the X-ray is dominated by the late-prompt emission \citep{nappo17}.}.
The fits were run \textit{without radio data}, and at the last MCMC iteration, the parameters of each individual walker were used to create a light curve (as the fitting is run with 500 walkers, this means 500 light curves per burst). The resulting distribution of light curves at radio frequencies was then compared to the radio data. If this distribution includes light curves that significantly under-predict the observed radio fluxes at the corresponding epochs (disregarding the \textit{shape} of the light curve), then the two-component scenario may be able to explain the light curve. If this is not the case, then an additional emission source would likely over-predict the radio fluxes. We include objects where the modeling code did not produce a good radio fit (GRBs 070125, 090313, 130907A, 140311A and 160625B), as this may have been due to the shape and peak time of the light curve instead of the general flux level. Two examples of the results of these tests are shown in Figure \ref{fig:2comp_test}.

For GRB 990510, which is consistent with the standard model regardless of the CBM type, the wind fits are able to reach a factor of $\sim20$ below the observed fluxes, while the ISM fits do so by a factor of $\sim2$. The radio fluxes of GRB 050820A are never significantly under-predicted in our ISM fits, but are under-predicted by a factor of $>10$ in wind fits -- the situation is very similar for GRB 140311A. Both of these seem to require an ISM-type CBM based on their optical and X-ray light curves, though. The radio fluxes of GRBs 090313 and 160625B are over-predicted in all of the test fits. On the other hand, the radio light curves of GRBs 070125 and 130907A are under-predicted in most test fits, and often stay below the observed flux level the entire time. This suggests that the two-component scenario may only be relevant for certain GRBs, or requires a mechanism to suppress the radio emission from the narrow jet; such a mechanism may be required in any case to produce radio quiet GRBs \citep{hancock13,ronning17}. However, the lack of radio quiet bursts with high $E_{iso}$ would then also need to be explained. 

Furthermore, in the two-component model one would naively expect to observe some GRBs whose radio emission clearly exceeds what the standard model predicts based on higher-frequency emission. We do, in fact, find one such object: GRB 130907A. We can only obtain a good fit to the X-ray and optical when ignoring the radio data, in which case the radio flux is under-predicted. This suggests that at least in the case of this particular GRB the scenario may very well be at work \citep[as suggested by][]{veres15}, especially since its radio decline is consistent with pre-break when $p\approx2$. However, as noted above, several of the afterglows seem to exhibit a steepening light curve simultaneously with the jet break, even if the following decline does not match asymptotic predictions. This argues against the presence of a secondary component, not associated with the forward shock of the main jet, dominating the radio emission at that time. 

\subsubsection{Reverse shock with energy injection}

It is possible to further test the possibility of a reverse shock and late-time energy injection suggested by \citet{pk04}. This is, however, a toy model that does not capture the full behavior of a reverse shock and its light curve under energy injection. In this scenario, the fireball energy $E \propto t^e$ and the mass of ejecta energized by the reverse shock $M (>\Gamma) \propto \Gamma^{-q}$. The decline of the radio light curve of the reverse shock is
\begin{equation}
-\alpha_\mathrm{RS} = \frac{3 - e}{2} + \frac{6 - k - e(2-k)}{4(4-k)}(p - 1) - \frac{3 - e - k}{2(4 - k)}q~.
\end{equation}
To use this equation to obtain $q$, one must first find $e$ using a higher frequency post-jet-break (forward shock) light curve. Without lateral expansion this is
\begin{equation}
-\alpha = \frac{3 - e}{4}p - \frac{e(16 - 5k) - k}{4(4-k)}~(\nu < \nu_c)
\end{equation}
or
\begin{equation}
-\alpha = \frac{3 - e}{4}p - \frac{e(6 - k) + k - 2)}{2(4-k)} ~(\nu_c < \nu)~,
\end{equation}
and with lateral expansion
\begin{equation}
-\alpha = p - \frac{p + 2}{3}e.
\end{equation}
These equations \citep[based on Eqs. 42, 43 and 30 in][]{pk04} assume that $\nu_{m}(\mathrm{RS}) < \nu_\mathrm{radio}$. We also check for the possibility that $\nu_{a}(\mathrm{RS}) < \nu_\mathrm{radio} < \nu_{m}(\mathrm{RS})$, in which case, from their Eq. 41, we can obtain
\begin{equation}
\alpha_{\mathrm{RS}} = \frac{(3q - 10)(3 - e - k) - 2k(1+e)}{6(4-k)}.
\end{equation}
In this scenario, $p$ must be determined using the spectrum. 

We estimate $e$ and $q$ for those GRBs in our sample where this test is possible, i.e. when an X-ray spectral index (from the references in Table \ref{tab:sample}), a post-jet-break light curve and a radio power law fit are available, and where a secondary component was not deemed implausible in Section \ref{sec:2comp}. The results are presented in Table \ref{tab:eqtab}. Problematic values are found for GRBs 051022 ($e<0$ implies energy loss, not injection, in an adiabatic jet) and 160509A ($e\approx0$ implies no injection). For GRB 130907A, the reverse shock and energy injection scenario may be in effect if $\nu_\mathrm{radio} < \nu_{m}(\mathrm{RS})$ and $k=2$, while in ISM scenarios $e<0$.

\begin{table*}
\centering
\begin{small}
\caption{Estimates of $e$ and $q$ parameters in the reverse shock and energy injection scenario.}
\begin{tabular}{lcc}
\hline
GRB & $\nu_{m}(\mathrm{RS}) < \nu_\mathrm{radio}$ & $\nu_\mathrm{radio} < \nu_{m}(\mathrm{RS})$ \\
 \hline
 051022 & $e\approx-0.4$; $q\approx4.0$ (ISM) or $q\approx-0.8$ (wind) & $e\approx-0.4$; $q\approx2.2$ (ISM) or $q\approx2.6$ (wind)\\
 070125 & $e\approx0.1$, $q\approx3.6$ & $e\approx0.1$, $q\approx1.7$ \\
 110709B & $e\approx0.2$, $q\approx3.6$ & $e\approx0.2$, $q\approx1.3$ \\
 130907A & $e\approx-0.3$, $q\approx3.6$ (ISM) or $e\approx0.1$, $q\approx-2.7$ (wind) & $e\approx-0.3$, $q\approx1.5$ (ISM) or $e\approx0.14$, $q\approx1.8$ (wind)\\
 140903A & $e\approx0.1$, $q\approx3.7$ & $e\approx0.1$, $q\approx1.6$ \\
 160509A & $e\approx0$, $q\approx2.8$ & $e\approx0$, $q\approx0.9$ \\
 \hline
\end{tabular}
\label{tab:eqtab}
\end{small}
\end{table*}

\subsection{Evolving microphysics}

Some GRBs have been reported to require time-evolving microphysics ($p$, $\epsilon_e$ or $\epsilon_B$); these include e.g. GRBs 091127 \citep{filgas11} and 190114C \citep{misra20}. In these GRBs time-evolving $\epsilon_B$ and/or $\epsilon_e$ were required to explain, respectively, the anomalous evolution of $\nu_c$ and a discrepancy between $\alpha$ and $\beta$. A related issue was brought up by \citet{gompertz18}, who noted numerous discrepancies between $\alpha$ and $\beta$ requiring an intrinsic scatter of $\sigma_p \approx 0.25$ on values of $p$ -- one suggested cause was evolving microphysics. The functional dependence of these parameters on time is not well constrained by theory, but the evolution of break frequencies and the slope of the light curve are changed relative to the values in Table \ref{tab:stdmodel}, possibly causing the light curve slopes seen in the sample. Signs of evolving microphysics could potentially be seen in some objects as, for example, a gradually changing X-ray photon index, which could be caused by an evolving $p$. GRB 130907A did exhibit a gradually rising X-ray photon index, but reached $\Gamma_X \approx 3.0$, which would require $p\approx 4$ if $\nu_m, \nu_c < \nu_X$; \citet{margutti15} ascribe this 'super-soft' X-ray spectrum instead to interaction with mass lost by the GRB progenitor. Such gradual $\Gamma_X$ evolution is not seen in other bursts in our sample. Evolving $\epsilon_e$ or $\epsilon_B$ could, on the other hand, cause anomalous break frequency evolution -- signs of this may be seen in GRB 171010A.

A proper investigation of this possibility is outside the scope of this paper, but we can briefly examine some examples. We use equations 3 and 5 in \citet{misra20} \citep[based on][]{wg99} and the fact that $f_\nu \propto f_m (\frac{\nu}{\nu_m})^{\frac{1}{3}}$ below $\nu_m$ and $f_\nu \propto f_m (\frac{\nu_c}{\nu_m})^{\frac{1-p}{2}} (\frac{\nu}{\nu_c})^{-\frac{p}{2}}$ above $\nu_c$ (assuming regime 1). We also assume, like them, that $p$ remains constant, while $\epsilon_e \propto t^\iota$ and $\epsilon_B \propto t^\lambda$. Then the pre-jet-break slope of the light curve below $\nu_m$ is $\alpha_\mathrm{ISM} = \frac{1}{6}(2\lambda - 4\iota + 3)$ or $\alpha_\mathrm{wind} = \frac{1}{6}(2\lambda - 4\iota)$.
Above $\nu_c$, we get $\alpha = \frac{1}{4}[p(\lambda-3) + 2(1-\lambda)] + \iota(p-1)$. We now look at a few GRBs where the X-ray light curve suggests no lateral expansion, in which case the post-break slope changes by $-3/4$ in ISM or by $-1/2$ in wind. For GRB 171010A, the seemingly pre-break $t^{-1.1}$ radio light curve and the $t^{-2.0}$ X-ray light curve can both be reproduced if $p \approx 2.3$, $\iota \approx 0.1$ and $\lambda \approx -2.3$; for GRB 160509A we can similarly estimate $p \approx 2.2$, $\iota \approx 0$ and $\lambda \approx -1.9$, and for GRB 110709B, $p \approx 2.2$, $\iota \approx 0.3$ and $\lambda \approx -0.7$. As a comparison, \citet{misra20} obtain $\iota \approx -0.4$ and $\lambda \approx 0.1$ for GRB 190114C after a more robust analysis, while \citet{filgas11} suggest $\iota = 0$, $\lambda = 0.49$ for GRB 091127. Thus there is a range of obtained parameters, both within this sample and between it and other results. Such variation does not rule out evolving microphysics, but does present another requirement for this potential solution.

\subsection{Other considerations}

It is noteworthy that some of the 'well-behaving' radio light curves (the aforementioned GRBs 990510, 050820A, 051022, and 110709B) may in fact be coincidental -- in these cases we only have one radio frequency with enough points for power-law fitting, and some other GRBs in the sample exhibit a radio SED that our model fits cannot reproduce. As a cautionary example, GRB 070125 would appear consistent with the standard model if the 95 GHz data and late 22.5 GHz points were not observed. This means that one must remain careful in these cases. The same applies to the objects in Group 2 with no X-ray data. GRBs 051022, 110709B, 140903A and 160509A, all consistent with the model, also suffer from a lack of optical constraints. Out of the eight GRBs with good optical and X-ray coverage mentioned above, we find three that fit the standard model. Those bursts where the slope is consistent with pre-break expectations may also be inconsistent with them at other frequencies. A further complication is that, in some cases, the radio spectrum indicates that the decline compatible with $\nu_m < \nu$ sets in even before the $\nu_m$ passage, or the shape of the spectrum stays flat. 

In addition to the two-component scenario mentioned above, other factors not included in our code can also have an effect on the light curve. A counter-jet was suggested as a way to explain the radio flux of GRB 980703 at $\sim1000$~d by \citet{perley17}, although the shape of the light curve is not conclusive. Counter-jets in general should peak roughly at $2(1+z)t_{NR}$, where $t_{NR}$ is the beginning of the non-relativistic phase \citep{zhangmac09}, and at the peak their flux should be $\sim6$ times the flux of the approaching jet. Most of our model fits place $t_{NR}$ after the last radio observations; often absence of a non-relativistic transition in the X-ray light curve constrains $t_{NR}$ to be late enough that the counter-jet cannot influence the radio light curve; and a late-time bump is usually not seen in the light curve. However, a counter-jet may conceivably be affecting GRB 070125. 

Apart from a rough test for whether our best fits may be affected by the Klein-Nishina limit \citep{nakar09}, described in Appendix \ref{sec:app2}, we have not included these effects in our code, as they are mostly ignored in GRB modeling efforts. These effects are not seen in the radio light curves, but a full inclusion of Klein-Nishina effects might result in some new sets of parameters being compatible with the X-ray data -- parameters that our code now excludes.

We also note that in nine cases (italicized in Table \ref{tab:models}) our code is unable to distinguish between a wind and an ISM density profile using the data available to us; both models result in almost equal $\chi^2$. Even for the other GRBs, unless only one of the models is consistent with the data (as with GRBs 050820A and 140903A), the other can typically not be excluded based on the model fits. Other observational signatures, such as the difference in light curve power laws above and below $\nu_m$, can be used to make the distinction in some cases such as GRB 990510, where the light curve indices favor the ISM profile, but interestingly enough, the wind model results in a slightly better fit. In reality,  the density profile may also not be as simple as is commonly assumed, particularly if eruptive mass loss dominates in the late stages of the star's evolution.

Finally we point out that an object outside our sample, GRB 170817A, eventually exhibited a $t^{-p}$ decline in the radio \citep{mooley18, lamb19, hajela19}. This particular afterglow requires a structured jet model with $p = 2.15^{+0.01}_{-0.02}$ and a relatively large observing angle of $30.4^{+4.0}_{-3.4}$~deg; because of the latter, the post-break decline was seen only when nearly non-relativistic. Its observation was made possible by a long follow-up, prompted by the gravitational waves and kilonova accompanying the event \citep[e.g.][]{abbott17,kilpatrick17,tanvir17}. A structured jet is expected to delay and soften a jet break \citep{kg03} and steepen the pre-break light curve \citep[when the structure function is a power law;][]{p05}. There should, however, be no steepening effect on post-break slopes. We also note that the objects in \citet{p05} where the top-hat jet model did not fit the data well were not fit well by the structured jet model either, and the fit was not improved in 4 out of 10 cases.

\section{Conclusions} \label{sec:concl}

We have examined a sample of 21 GRB afterglows with evidence of a jet break, but with otherwise varying properties. The sample is divided into two groups: Group 1 consists of a representative sample of 12 GRBs from the \emph{Swift} catalog, while Group 2 adds in 9 GRBs found through a literature search -- these either predate \emph{Swift} or do not match all of the search criteria. We have used both single/broken power law fits to the light curves and a fitting code based on the standard model \citep{mrees99,rhoads99,sariesin01,granotsari02,vdhthesis}.

In most cases, while conventional fireball/jet theory can provide a good fit to the X-ray (and optical) light curve, it does a worse job with their radio light curves and/or SEDs. The decline of the radio light curve is mostly well described by a single power law inconsistent with predictions of the post-break decline. Sometimes (eight cases out of 21) the single power law is consistent with theoretical predictions for the \textit{pre}-break decline, however, even up to several tens of times later than the jet break time. 

According to our model fitting code, the standard model is able to account for the observed behavior of roughly half of all the examined GRBs, and a fraction of $0.42_{-0.17}^{+0.19}$ of the more representative Group 1. However, even in these cases the radio data mostly do not show a power-law decline consistent with post-break predictions. Even when the standard model fits the data anyway, in several cases it requires a steepening to $\sim t^{-2}$ soon after the last radio observations. Furthermore, three of these bursts have little to no optical data to constrain the fits. While GRB 120326A is an example of a GRB that eventually exhibited this steepening, the prediction highlights the importance of long radio follow-up of individual GRBs in order to confirm or exclude it. We also note that in many cases, the difference in $\chi^2$ between ISM and wind models is very small, making it difficult to firmly establish the density profile using the available data.

All in all, in individual cases, the model fit in the radio may seem adequate, especially considering the typical errors and often small numbers of data points per band in the literature, but looking at a larger sample of events, a pattern emerges. A single power law fit to the radio, mostly inconsistent with asymptotic power law predictions, tends to describe the light curve at least as well as model fits based on the standard jet model. We investigate the possibility of a secondary component (a wider jet or reverse shock emission with energy injection) dominating the late-time radio emission. This addition may explain a part of the anomalous afterglows, but in some cases requires another effect to suppress the emission from the forward shock of the narrow jet, and is disfavored in the cases where some light-curve steepening is seen simultaneously with the jet break. Evolving microphysics parameters may reproduce at least some of the observed post-jet-break light curves.

As Group 1 of our sample includes all GRBs in the \emph{Swift} catalog with X-ray jet breaks and published radio data, it is both representative and contains GRBs with a range of properties. Thus the large fraction of objects whose behavior cannot be reproduced by our model fits, along with the scarcity of unambiguous cases of objects that behave as predicted by the standard model (and the lack of constraints associated with them), is telling. Along with the radio-quiet GRB population \citep{hancock13, ronning17, ronning18}, this highlights our lack of understanding of this part of the GRB afterglow spectrum and, quite possibly, the relevant physics. This problem may be better investigated in the future through radio and millimeter follow-up programs which continue until long after an observed X-ray or optical break. Until then, the physical parameters determined for GRBs may contain systematic errors. \\

\acknowledgments

We thank the two anonymous referees for many comments and suggestions that prompted extensive changes to this paper, improving it considerably. We also thank Dale Frail, Alexander van der Horst, Asaf Pe'er and Chryssa Kouveliotou for helpful feedback and discussion.

Based on observations made with the NASA/ESA \textit{Hubble Space Telescope} (programme GO 14353, PI Fruchter), obtained through the data archive at the Space Telescope Science Institute (STScI). STScI is operated by the Association of Universities for Research in Astronomy, Inc. under NASA contract NAS 5-26555. Support for this work was also provided by the National Aeronautics and Space Administration through Chandra Award Number 17500753, PI Fruchter, issued by the Chandra X-ray Center, which is operated by the Smithsonian Astrophysical Observatory for and on behalf of the National Aeronautics Space Administration under contract NAS8-03060. This work made use of data supplied by the UK Swift Science Data Centre at the University of Leicester.

%

\vspace{5mm}





\appendix

\section{Lateral expansion and non-relativistic phase below $\nu_{\MakeLowercase{ac}}$}
\label{sec:app1}

Below $\nu_{ac}$, the behavior of the light curve is given in \citet{granotsari02} for the pre-break and edge-effect break cases. However, the source of most of the light curve power-law segments in the cases of lateral expansion and non-relativistic phase, \citet{vdhthesis}, ignores $\nu_{ac}$ entirely. Thus, for completeness, we use Eq. 5 in \citet{gps00} to get
\begin{equation}
\nu_{ac}(t) = \Big(\frac{6m_e^{12} c^{29} \Gamma^5 n'^{11}}{\pi^5 \epsilon_B^2 \epsilon_e^8 q_e^4 e'^{10}}\Big)^{1/5} \propto \Big(\frac{\Gamma(t)^5 n'(t)^{11}}{e'(t)^{10}}\Big)^{1/5},
\end{equation}
where $\Gamma(t) \propto t^{-1/2}$ in the jet-spreading phase and $\Gamma(t) \approx 1$ in the non-relativistic phase; $n'(t) \propto \Gamma(t) n(t)$ and $e'(t) \propto \Gamma(t)^2 n(t)$. Therefore in the jet spreading phase, where the jet radius $R\approx \mathrm{const}$ \citep[e.g.][]{kumarzhang15}, we obtain $\nu_{ac} \propto \Gamma(t)^{-4/5} \propto t^{4/10}$ in both ISM and wind cases.
From 
\begin{equation}
F_{\nu}(\nu<\nu_{ac}<\nu_a<\nu_c<\nu_m) = F_{\nu,\mathrm{max}} \Big(\frac{\nu_a}{\nu_c}\Big)^{1/3} \Big(\frac{\nu_{ac}}{\nu_a}\Big)^{11/8} \Big(\frac{\nu}{\nu_{ac}}\Big)^{2}~,
\end{equation}
\begin{equation}
F_{\nu}(\nu_{ac}<\nu<\nu_a<\nu_c<\nu_m) = F_{\nu,\mathrm{max}} \Big(\frac{\nu_a}{\nu_c}\Big)^{1/3} \Big(\frac{\nu}{\nu_a}\Big)^{11/8}
\end{equation}
and using Table 2.10 of \citet[][]{vdhthesis} for the behavior of $F_{\nu,\mathrm{max}}$, $\nu_c$ and $\nu_a$, we obtain $F_{\nu}(\nu<\nu_{ac}<\nu_a<\nu_c<\nu_m) \propto t^0$ and $F_{\nu}(\nu_{ac}<\nu<\nu_a<\nu_c<\nu_m) \propto t^{1/4}$.
In the non-relativistic phase and with ISM-type CBM, both $\Gamma$ and $n$ are roughly constant and hence $\nu_{ac} \propto t^0$. From Eqs. A2 and A3 and Table 2.11 of \citet[][]{vdhthesis} it then follows that both $F_{\nu}(\nu<\nu_{ac}<\nu_a<\nu_c<\nu_m)$ and $F_{\nu}(\nu_{ac}<\nu<\nu_a<\nu_c<\nu_m)$ evolve as $t^{7/8}$. In a wind-type CBM, we have $n(t) \propto R(t)^{-2} \propto t^{-4/3}$, and thus, $\nu_{ac} \propto t^{-4/15}$. Using the same table, we thus obtain $F_{\nu}(\nu<\nu_{ac}<\nu_a<\nu_c<\nu_m) \propto t^{-121/72}$ and $F_{\nu}(\nu_{ac}<\nu<\nu_a<\nu_c<\nu_m) \propto t^{-133/72}$.

\section{The possibility of Klein-Nishina effects}
\label{sec:app2}

As in some cases our modeling code predicts that IC effects \citep{sariesin01} are important, we also test whether the best-fit model of each afterglow is affected by Klein-Nishina (KN) effects \citep{nakar09}. As the energies of photons become comparable to the electron rest mass at the KN limit, the scattering cross section changes and electron cooling is slowed, hardening the spectrum at high energies ($\nu > \nu_c$); this can happen in some GRB afterglows, but most afterglows are only expected to be weakly affected.

Our code, based on the standard afterglow theory widely used by the GRB community, does not include the full analytical treatment of KN effects on the spectrum as described in \citet{nakar09}. However, we can test each best-fit model for significant KN suppression of IC cooling. This is done using their Eq. 64 for the maximum value of the $Y$ parameter:
\begin{equation}
Y_{max}(\gamma_c) \approx 10 e^{\frac{27(3.2-p)^2 - 17}{(p+2)(4-p)}} \epsilon_{e,-1}^{\frac{5(p-1)}{p+2}} E_{53}^{1/2} n^{\frac{8-p}{2(p+2)}} t_d^{-\frac{5(p-2)}{2(p+2)}}~.
\end{equation}
This equation can be used directly for ISM-type CBM, but in the wind case the density of the medium the shock moves into evolves as $n(t) = (3\times10^{35}~\mathrm{cm}^{-1}) A_* R^{-2}$, the radius as $R \approx 2ct\Gamma^{2}$ and the bulk Lorentz factor as $\Gamma \approx (1/\theta_j) (t/t_j)^{-1/4}$ \citep[c.f.][]{kumarzhang15}. If $Y_{max}(\gamma_c)$ at $t=1$~d is below the value of $Y$ in regime 5 (i.e. the highest $Y$ predicted by the model), and the IC-corrected $\nu_c$ is below the highest observed frequency, we conclude that KN effects are affecting the best fit.

\begin{figure*}
\begin{minipage}{\linewidth}
\centering
\includegraphics[width=0.45\columnwidth]{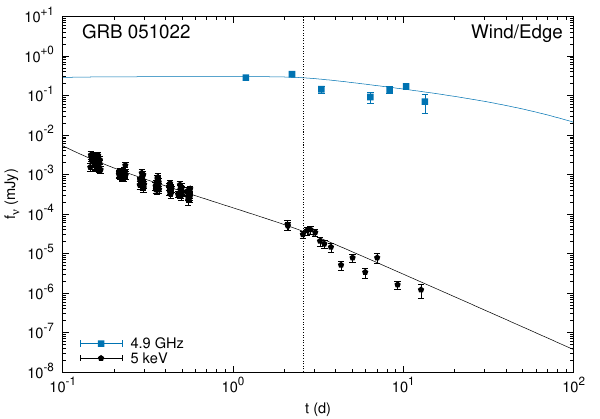} \includegraphics[width=0.45\columnwidth]{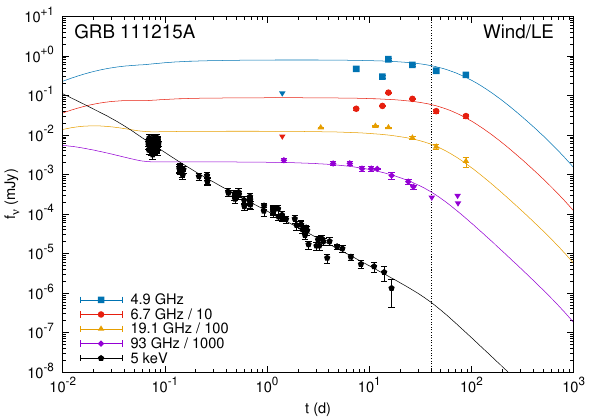} \\
\includegraphics[width=0.45\columnwidth]{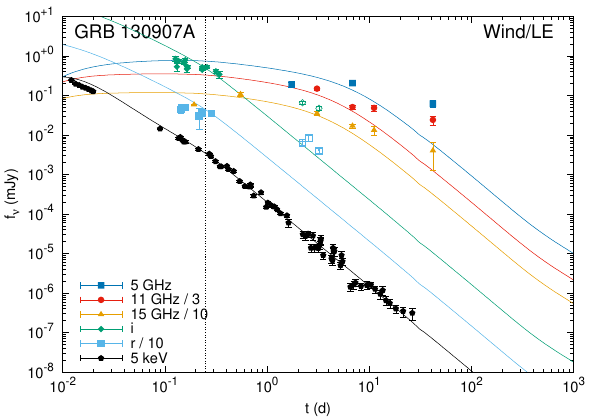} \includegraphics[width=0.45\columnwidth]{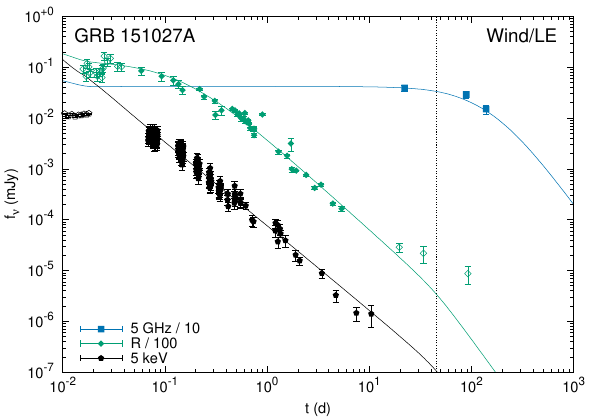}
\caption{The best fits from the test for Klein-Nishina effects, run with a limitation on the maximum value of the $Y$ parameter from Eq. A1.}
\label{fig:knfig}
\end{minipage}
\end{figure*}

In most cases, we find $Y_{max}(\gamma_c,1~\mathrm{d}) \gg Y$, or a $\nu_c$ too high to have been observed, and the best fit is therefore unaffected. However, this is not the case for GRBs 051022, 111215A, 130907A and 151027A. For these GRBs, as a rough approximation, we have run a version of our code that incorporates the constraint $Y(t) \leq Y_{max}(\gamma_c,t)$, and assume that KN effects on IC \textit{emission} are confined to higher frequencies than observed. As a result of this, the quality of the best fit remains similar in all cases except GRB 130907A, where the fit at X-rays is improved but the radio behavior is still unaccounted for. Our conclusions therefore do not change. The resulting parameters of these fits are given in Table \ref{tab:kntab} and the best-fit light curves presented in Figure \ref{fig:knfig}.

\begin{table*}
\centering
    \begin{tabular}{lcccccccc}
       GRB & Model & $E_\mathrm{K,iso}$ & $p$ & $n_0$ (cm$^{-3}$) & $\epsilon_e$ & $\epsilon_B$ & $\theta_j$ & $A_V$ \\
       &  & ($10^{52}$ erg) & & or $A_{*}$ & &  & (rad) & (mag) \\
       \hline
        051022 & Wind/Edge & $13.4_{-3.4}^{+8.2}$ & $2.37_{-0.06}^{+0.11}$ & $0.10_{-0.06}^{+0.10}$ & $0.21_{-0.05}^{+0.04}$ & $1.9_{-1.6}^{+13.5} \times10^{-3}$ & $0.06_{-0.01}^{+0.02}$ & ...\footnote{For dark GRBs (GRBs 051022 and 111215A), optical extinction is not relevant and was fixed at zero.} \\
        111215A & Wind/LE & $219_{-23}^{+22}$ & $2.45\pm0.02$ & $0.12\pm0.02$ & $0.06\pm0.01$ & $0.02\pm0.01$ & $0.06\pm0.01$ & ... \\
        130907A & Wind/LE & $421_{-55}^{+77}$ & $2.12\pm0.01$ & $0.31_{-0.04}^{+0.05}$ & $0.17\pm0.02$ & $6.9_{-1.6}^{+2.4} \times 10^{-5}$ & $0.02\pm0.01$ & 1.3\footnote{For GRB 130907A, extinction was fixed at $A_V = 1.3$ as per \citet{veres15} (see text).}  \\
        151027A & Wind/LE & $125_{-50}^{+41}$ & $2.78_{-0.03}^{+0.02}$ & $5.6_{-2.4}^{+9.3}\times10^{-3}$ & $0.04_{-0.01}^{+0.03}$ & $0.31_{-0.23}^{+0.38}$ & $0.03_{-0.01}^{+0.03}$ & $0.5\pm0.1$ \\
        \hline
    \end{tabular}
    \caption{MCMC fit parameters and their uncertainties for each GRB tested for KN effects, assuming $Y(t) \leq Y_{max}(\gamma_c,t)$.}
    \label{tab:kntab}
\end{table*}




\bibliographystyle{aasjournal}
 \bibliography{references}





\end{document}